\documentclass[12pt]{article}
\usepackage[top=2.5cm, left=2.5cm, bottom=2.5cm, right=2.5cm]{geometry}
\usepackage[english]{babel}
\usepackage[square,numbers]{natbib}
\usepackage{mathptmx}
\usepackage{graphicx}
\usepackage{subcaption}
\usepackage{siunitx} 
\usepackage{amsmath,amsthm,amssymb,mathtools}
\usepackage[affil-it]{authblk}
\usepackage{xurl}
\usepackage{indentfirst}
\usepackage{microtype}
\usepackage[colorlinks = true, linkcolor = black, citecolor=black, urlcolor=blue, filecolor = black, bookmarks=false,hypertexnames=true]{hyperref}
\graphicspath{{./}}

\title{Analyzing Errors in Controlled Turret System}

\author[1]{Matthew Karlson}
\author[1]{Heng Ban \thanks{Corresponding Author. \textit{Email Address: heng.ban@pitt.edu}}}
\author[1]{Daniel G. Cole}
\author[2]{Mai Abdelhakim}
\author[3]{Jennifer Forsythe}
\author[3]{John T. Fitzgibbons}

\affil[1]{Department of Mechanical Engineering and Materials Science \protect\\ University of Pittsburgh, Pittsburgh, PA 15260 \vspace{0.5em}}
\affil[2]{Department of Electrical and Computer Engineering \protect\\ University of Pittsburgh, Pittsburgh, PA 15260
\vspace{0.5em}}
\affil[3]{DEVCOM Analysis Center \protect\\ Aberdeen Proving Ground, MD 21005}

\date{}

\begin{document}

\maketitle

\begin{abstract}
	The purpose of this paper is to characterize aiming errors in controlled weapon systems given target location as input. To achieve this objective, we analyze the accuracy of a controlled weapon system model for stationary and moving targets under different error sources and firing times. First, we develop a mathematical model of a gun turret and use it to design two controllers, a Proportional-Integral-Derivative controller and a Model Predictive controller, which accept the target location input and move the turret to the centroid of the target in simulations. For stationary targets, we analyze the impact of errors in estimating the system's parameters and uncertainty in the aimpoint measurement. Our results indicate that turret movement is more sensitive to errors in the moment of inertia than the damping coefficient, which could lead to incorrect simulations of controlled turret system accuracy. The results also support the hypothesis that turret movement errors are larger over longer distances of gun turret movement and, assuming no time constraints, accuracy improves the longer one waits to fire; though this may not always be practical in a combat scenario. Additionally, we demonstrate that the integral control component is needed for high accuracy in moving target scenarios.
\end{abstract}

\clearpage
\section{Introduction}\label{introduction}
This paper seeks to characterize aiming errors in weapon systems that use controllers to move the turret and gun given a target location as input. Next-generation weapons systems will rely on increasingly sophisticated fire control methods to include aiming. The need to analyze errors in the motion controller is emphasized by the growing research of using artificial intelligence (AI) to recognize targets and pass their location to the weapon systems.

Most research on improving weapon system accuracy has focused predominately on human lay error. Generally, this type of lay error is assumed to be a random source of error that is normally distributed over the target's impact area \citep{groves1963method,helgert1969statistical,strohm2013introduction}. This is due to its dependency on various factors such as unsteady motion of the gunner's hand, the firing position of the gunner, the stress response induced in combat, the level of training, and target range. Consequently, by considering human lay error to be part of a larger error budget encompassing all statistically relevant sources of ballistic delivery error, theoretical predictions of the probability of a hit have been derived \citep{groves1963method,helgert1969statistical}.

Researchers in the academic and military communities have conducted field experiments to test these predictions. This is used to develop error budgets, a statistical categorization of system accuracy components, as a function of target range for different classes of firearms, and to examine the factors that significantly influence human lay error. \citet{Weaver1990system} and \citet{wahlde1999sniper} develop error budgets for conventional military-grade general-purpose rifles and sniper rifles based on experimental data and theoretical estimates of the distributions of each source of delivery error, including human lay error. \citet{torre1991effects} develop a methodology for inducing stress in soldiers and measuring their shooting performance in experimental trials. \citet{Dwyer2003improved} examine the effects of unsteady motion of the hand on a soldier's aim in different firing positions and study their impacts on the probability of a hit against static and moving targets at different ranges. Similarly, \citet{Corriveau2019firingposition} evaluate lay error in experimental trials with soldiers under induced stress conditions and in different firing positions against static targets at varying ranges. The results from these experiments show that human lay error has significant impacts on weapon accuracy beginning at a range of 300 \si{\meter}, reducing the probability of a hit over longer ranges. Overall, these experimental results support the assumption that human lay error is a random source of error that is normally distributed over the target's impact area \citep{Corriveau2019firingposition,Dwyer2003improved}.

Due to the limitations of human aiming ability, various control approaches have been developed to improve firing accuracy by using a feedback controller to assist in turret movement. For purposes of research, we assume a target location is provided and are focused on the error in moving the turret from an initial location to the location of least error. The more common methods for this objective are based on Proportional-Integral-Derivative (PID) control \citep{carlstedt2021modelling,idris-hudha-ASCC2015,Kuswadi2016,lyth2021modelling,Nasyir2014}. Other recent approaches are based on Adaptive Robust Control (ARC) \citep{Ma2022adaptive,Yuan2021precision,Yuan2024bidirectional}, Sliding Mode Control (SMC) \citep{rahmat-et-al-AIMT2016,Xia2016modeling}, Model Predictive Control (MPC) \citep{Kumar2009study}, and methods that synthesize controllers using AI  \citep{Anwar2017deep,Ceceloglu2019modeling}. The advantages of using feedback control for aiming over manual aiming are, but not limited to, a reduced sensitivity of the turret system to extraneous vibrations of the gun barrel and to the effects of the stress response that is naturally inherent in humans. While considerable progress has been made in control research for turret systems, the work done has centered only on demonstrating the viability of a particular control approach rather than statistically analyzing the accuracy of the weapon system under feedback control.

In this paper, we analyze the aiming error distributions of a controlled gun turret system given target location as input through numerical simulations. We develop a mathematical model of the gun turret based on Newton's laws. Based on the developed model for the physical system, we design two controllers, PID and MPC, to simulate controlled aiming at static targets. In numerical experiments, we study the aiming error distribution under several scenarios of controlled gun turret movement. First, we conduct a sensitivity analysis to analyze the impact of estimation errors in model parameters on the error distribution. We then perform an experiment that examines the effects of uncertainty in the aimpoint measurement on gun turret accuracy. Next, we analyze the dependency of aiming accuracy on when one chooses to fire at a target using the error data from the sensitivity analysis and the uncertainty experiment. Continuing with uncertainty analysis, we perform another experiment wherein we model the process of measuring the aimpoint and quantify the uncertainty added to the aiming error. In this experiment, we calculate the error distribution statistics analytically and compare the result to numerical estimates of the statistics from simulation data. In addition to a stationary target, we consider moving targets. We design two variants of PID controllers for aiming at moving targets and analyze the performance of the control system. 
\section{Methods}
\label{methods}

\subsection{Gun Turret Model}
\label{gunturretmodel}

\begin{figure}
    \centering    
    \includegraphics[scale=0.3]{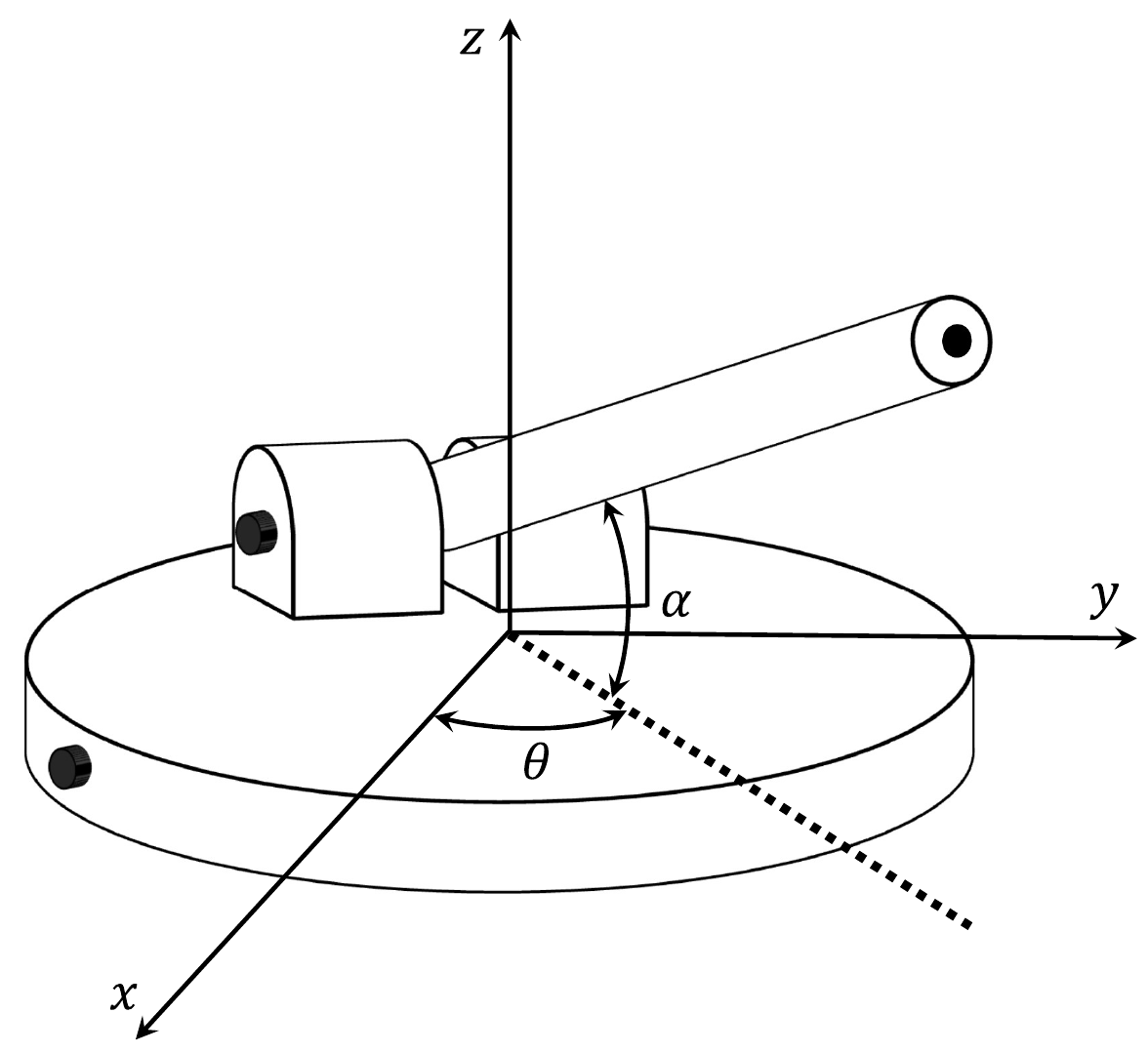}
    \caption{Illustration of the gun turret system. The frame $\{x,y,z\}$ is fixed to the ground and the $z$-axis is coincident with the center of mass of the platform.}
    \label{gun_turret_model}
\end{figure}

The gun turret we model for our controller-aided aiming simulations consists of a platform and gun barrel as its two main components. The platform and gun barrel are modeled as two rigid bodies with the pivot point of the gun barrel coinciding with the center of mass of the platform as shown in Figure \ref{gun_turret_model}. We ignore coupling effects due to Coriolis and centripetal forces. This is because we assume targets are either static or moving at small angular speeds ($\le \SI{10}{\degree/\second}$). Under these assumptions, the terms due to Coriolis and centripetal forces do not contribute significantly to the dynamics because they are proportional to the products of the angular speeds of the platform and gun barrel. Additionally, we ignore external disturbances to the system and limit the elevation angle to \SI{30}{\degree}. The inertia of each motor is also assumed to be negligible and we neglect the forces to fire the weapon since we are studying only angular movement of the turret. We include damping torques due to bearing friction proportional to the angular speeds of each rigid body. The inputs are the motor torques $T_1$ driving the platform and $T_2$ driving the gun barrel; both are measured in \si{\newton\cdot\meter}. Referring to Figure \ref{gun_turret_model}, the outputs are $\theta$, the azimuth position of the platform from the positive $x$-axis and $\alpha$, the elevation position of the gun barrel from the dashed line. The outputs are measured in \si{\radian}.

The model parameters used in this study are shown in Table \ref{lineargunturretparametervalues}. The parameters are chosen in consideration of previous work on turret control systems \citep{carlstedt2021modelling,lyth2021modelling,Ma2022adaptive,Yuan2024bidirectional} and state-of-the-art main battle tanks such as the M1 Abrams \citep{m1abramsSpecs}.
The platform is modeled as a disk of radius $R$ and the gun barrel is modeled as a rod of length $L$. We estimate the moment of inertia of each body as
\begin{equation}
     J_1 = \frac{1}{2}m_1R^2 + J_2, \quad J_2 =  \frac{1}{3}m_2L^2.
     \label{moiequations}
\end{equation}

The moment of inertia of the platform $J_1$ in \eqref{moiequations} is estimated from
\begin{equation}
    J(\alpha) = \frac{1}{2}m_1R^2 + J_2\cos^2\alpha,
    \label{Jptrue}
\end{equation}
which assumes the pivot point of the gun barrel lies on the center of the platform. Since we limit the elevation to \SI{30}{\degree}, the error in the approximation of \eqref{Jptrue} by $J_1$ is about \SI{20}{\percent}. However, in a physical system, this error can be accounted for by designing the controller for a \SI{20}{\percent} larger moment of inertia.

The equations of motion of the gun turret are derived by summing the moments acting on the platform and gun barrel. The details are given in Section \ref{eqnsofmotion} in the Appendix. This leads to the following two equations for $\theta$ and $\alpha$:
\begin{align}
    J_1\ddot{\theta} &+ b_1\dot{\theta} = T_1, \label{thetaeqnofmotion}\\
    J_2\ddot{\alpha} &+ b_2\dot{\alpha} + \frac{1}{2}m_2gL\cos\alpha = T_2  \label{alphaeqnofmotion}.
\end{align}

\begin{table} 
    \centering
    \caption{Parameters of the gun turret model}
    \begin{tabular}{cll}
        \hline
        \textbf{Parameter} & \multicolumn{1}{c}{\textbf{Description}} & \textbf{Value} \\
         \hline
         $m_1$ & Mass of the platform & \num{8.67e3} [\si{\kilogram}] \\
         $m_2$ & Mass of the gun barrel & \num{4.97e3} [\si{\kilogram}] \\
         $b_1$ & Damping coefficient of the platform & \num{6.00e4} [\si{\newton\cdot\meter/\second}] \\
         $b_2$ & Damping coefficient of the gun barrel & \num{6.00e4} [\si{\newton\cdot\meter/\second}] \\
         $J_1$ & Moment of inertia of the platform & \num{7.99e4} [\si{\kilogram\cdot\meter^2}] \\
         $J_2$ & Moment of inertia of the gun barrel & \num{4.83e4} [\si{\kilogram\cdot\meter^2}] \\
         $R$ & Radius of the platform & \num{2.70} [\si{\meter}] \\
         $L$ & Length of the gun barrel & \num{5.40} [\si{\meter}] \\
         \hline
    \end{tabular}
    \label{lineargunturretparametervalues}
\end{table}

\subsection{System Transfer Functions}

The advantage of a linear model for the gun turret is that we can treat the azimuth and elevation variables independently in the control system design. From the equations of motion \eqref{thetaeqnofmotion} and \eqref{alphaeqnofmotion}, we can obtain two single-input-single-output transfer functions that can be used in the design of the PID controllers. The derivation is as follows.

First, since $\alpha \le \SI{30}{\degree}$, then $\cos\alpha \simeq 1$ from the small-angle approximation. Substituting this into \eqref{alphaeqnofmotion} gives
\begin{equation}
    J_2\ddot{\alpha} + b_2\dot{\alpha} + \frac{1}{2}m_2gL = T_2.
    \label{approxalphadiffeqn}
\end{equation}
Next, we assume aiming of the gun barrel begins from a point of static equilibrium, which allows us to drop the gravitational torque term $\frac{1}{2}m_2gL$ in \eqref{approxalphadiffeqn} since it is a constant disturbance to the system that will be attenuated by the controllers designed in this study. Moreover, the added error resulting from this input is a constant bias at a given firing time that can be estimated and compensated for by adjusting the gun barrel. Then by letting $u_1 = T_1$ and $u_2 = T_2$, some algebraic manipulation leads to
    \begin{align} 
    \ddot{\theta} + c_1\dot{\theta} &= A_1u_1 \label{thetadiffeqn},\\
    \ddot{\alpha} + c_2\dot{\alpha} &= A_2u_2,
    \label{alphadiffeqn}
    \end{align}
where
\begin{equation*}
     \begin{aligned}
         A_1 &= \frac{1}{J_1}, \quad A_2 = \frac{1}{J_2}, \\
         c_1 &= \frac{b_1}{J_1}, \quad c_2 = \frac{b_2}{J_2}.
     \end{aligned}
\end{equation*}

Since we assume the azimuth and elevation each start from a position of \SI{0}{\degree}, taking the Laplace transform on both sides of \eqref{thetadiffeqn} and \eqref{alphadiffeqn} gives
\begin{align}
    Y_1(s) &= \frac{A_1}{s(s+c_1)}U_1(s) \label{thetasdomaineqn}\\
    Y_2(s) &= \frac{A_2}{s(s+c_2)}U_2(s),
    \label{alphasdomaineqn}
\end{align}
where $s$ is the frequency variable with units \si{\radian/\second} and $Y_1(s)$, $Y_2(s)$, $U_1(s)$, and $U_2(s)$ are the Laplace transforms of $\theta$, $\alpha$, $u_1$, and $u_2$, respectively. 

The rational functions
\begin{align}
        G_1(s) &= \frac{A_1}{s(s+c_1)} \label{thetatf}\\
        G_2(s) &= \frac{A_2}{s(s+c_2)}.
        \label{alphatf}
\end{align}
on the right-hand side of \eqref{thetasdomaineqn} and \eqref{alphasdomaineqn} are the system transfer functions relating the input $u_1$ to the output $\theta$ and the input $u_2$ to the output $\alpha$.

\subsection{State-Space Representation}
\label{ssrepresentation}

By linearizing the equations of motion \eqref{thetaeqnofmotion} and \eqref{alphaeqnofmotion} about an equilibrium point, we can obtain a linear state-space representation of the gun turret system, which can be used in the design of the MPC controller. The linearization procedure is described in Section \ref{linearssmodelderivation} of the Appendix. We define the state variables to be
$x_1 = \theta$, $x_2 = \dot{\theta}$, $x_3 = \alpha$, and $x_4 = \dot{\alpha}$,
while the inputs are
$u_1 = T_1$ and $u_2 = T_2$,
and outputs are
$y_1 = x_1$ and $y_2 = x_3$.
The state vector is $\mathbf{x}=[x_1,x_2,x_3,x_4]^{\mathrm{T}}$ and the input vector is $\mathbf{u}=[u_1,u_2]^{\mathrm{T}}$. We choose the nominal input $\mathbf{u}_0 = [0,mgL/2]^{\mathrm{T}}$ and solve for the equilibrium point $\mathbf{x}_0=[0,0,0,0]^{\mathrm{T}}$. 

After performing the linearization about the equilibrium point given the nominal input, the result is the linear system
    \begin{align}
    \dot{\mathbf{x}}
    &= \begin{bmatrix}
        0 & 1 & 0 & 0 \\
        0 & -\frac{b_1}{J_1} & 0 & 0 \\
        0 & 0 & 0 & 1 \\
        0 & 0 & 0 & -\frac{b_2}{J_2}
    \end{bmatrix}
    \mathbf{x}
    +
    \begin{bmatrix}
     0 & 0 \\
     \frac{1}{J_1} & 0 \\
     0 & 0 \\
     0 & \frac{1}{J_2}
    \end{bmatrix}
    \mathbf{u}
    \label{xdotsystem} \\[1ex]
    \mathbf{y}
    &= 
    \begin{bmatrix}
    1 & 0 & 0 & 0 \\
    0 & 0 & 1 & 0
    \end{bmatrix}
    \mathbf{x}
    +
    \begin{bmatrix}
    0 & 0 \\
    0 & 0
    \end{bmatrix}
    \mathbf{u}.
    \label{ysystem}
    \end{align}
The system of equations \eqref{xdotsystem} and \eqref{ysystem} is the state-space representation of the gun turret for small deviations away from $\mathbf{x}_0$ and $\mathbf{u}_0$.

\subsection{PID Control}
\label{pidcontrol}
PID control is widely used in industry for its simple design and effectiveness in controlling plant processes. In control theory, a plant is the system encompassing the actuator and process wherein the process produces outputs in response to inputs received from the actuator. In the context of this work, the actuators are the motors driving the platform and gun barrel, and the process is the gun turret. A PID controller is a device that determines the appropriate actuation commands to achieve a desired output from the process. The actuation command is calculated from the PID control law, which in the time domain is
\begin{equation}  
    u(t) = K_Pe(t) + K_I\int e(t)\, dt + K_D \dot{e}(t).
    \label{pideqn}
\end{equation} 
In \eqref{pideqn}, $e(t) = r(t)-y(t)$ is the tracking error between the desired output $r(t)$ and system output $y(t)$, and $K_P$, $K_I$, and $K_D$ are gains, referred to as proportional gain, integral gain, and derivative gain, respectively. The gains should be turned to achieve the control objective. 

In this study, we use frequency response methods applied to the transfer functions in \eqref{thetatf} and \eqref{alphatf} to design two separate controllers for the platform and gun barrel. In this approach, we specify a bandwidth requirement to handle speed and determine an adequate phase margin (PM) for stability. These requirements are two conditions we want the controllers to satisfy to achieve the design objectives in Section \ref{controllerobjectives}.

The bandwidth $\omega_{bw}$ of the closed-loop system is defined as the frequency where the amplitude of the transfer function is approximately 0.707 \citep{FranklinCh6FreqResp}. The gain-crossover frequency $\omega_{gc}$, which is the frequency where the amplitude of the loop gain $L(j\omega) = C(j\omega)G(j\omega)$ is unity, can be used to estimate the bandwidth based on the following rule of thumb:
    $\omega_{gc}\le \omega_{bw} \le 2\omega_{gc}$
The PM measures the degree to which stability conditions are met. These are conditions on the magnitude and phase of the loop gain. The requirement for stability is $|L(j\omega)| < 1$ at $\angle G(j\omega) = -180^{\circ}$ for systems for which increasing the gain leads to instability.

For the platform azimuth motion, $\theta$, we design a lead controller which approximates pure PD control ($K_I=0$). The transfer function for a lead controller in the frequency domain is
\begin{equation}
    C(s) = K_P\left(\frac{T_Ds+1}{\gamma T_Ds+1}\right),
    \label{plead}
\end{equation}
where $0 < \gamma < 1$ and $T_d > 0$. The lead controller adds extra phase to the open-loop system needed to achieve the desired PM. More generally, lead control speeds up the transient response while providing some noise attenuation at higher frequencies. This controller is used in all experiments with static targets.

For the elevation, $\alpha$ of the gun barrel, we design a PI+lead controller which approximates pure PID control. The transfer function for a PI+lead controller in the frequency domain is
\begin{equation}
    C(s) = K_P\left(\frac{s+1/T_I}{s}\right)\left(\frac{T_Ds+1}{\gamma T_Ds+1}\right),
    \label{pilead}
\end{equation}
where $0 < \gamma < 1$, $T_D > 0$, and $T_I > 0$. The integral term reduces steady-state errors and adds disturbance rejection. This controller is used in all experiments in this work.

For moving targets, we design another PI+lead controller for the platform azimuth motion, $\theta$, and compare the performance with the lead controller.  The design procedure for both controllers is described in Sections \ref{leaddesign} and \ref{pileaddesign} in the Appendix. The controller parameters used in this study are listed in Table \ref{azimuthelevationcontrollerparameters}

\begin{table}
    \centering
     \caption{PID controller parameters for the gun turret system. The last two columns report the measured $\omega_{gc}$ and PM for each controller after the design procedure.}
    \begin{tabular}{cccccccc}
    \hline
    \textbf{Variable} & \textbf{Controller} & $K_P$ & $T_D$ [\si{\second}] & $T_I$ [\si{\second}] & $\gamma$ &  $\omega_{gc}$ [\si{\hertz}] & $\mathrm{PM}$ [\si{\degree}] \\
    \hline
      $\theta$ & lead & \num{2120360}. & \num{0.42} & - & \num{0.045} & \num{1.78} & \num{70.0}  \\
     $\theta$ & PI+lead$_{\theta}$ & \num{6507863} & \num{0.32} & \num{0.45} & \num{0.020} & \num{4.19} & \num{70.7} \\
    $\alpha$ & PI+lead$_{\alpha}$ & \num{1387706} & \num{0.45} & \num{0.78} & \num{0.030} & \num{2.06} & \num{70.3}
    \\
    \hline
    \end{tabular}
    \label{azimuthelevationcontrollerparameters}
\end{table}

\subsection{Model Predictive Control}
\label{mpccontrol}
Model Predictive Control (MPC) is a method of process control for dynamic systems. MPC uses a mathematical model of the process to make predictions of future plant outputs, denoted ``measured outputs" (MOs), with the aim of determining optimal control inputs over a prediction horizon, which is the number of intervals ahead of the current control interval. The control inputs, referred to as ``manipulated variables" (MVs), are calculated by using the MO predictions to solve a constrained optimization problem that minimizes a cost function based on control objectives \citep{Garcia1989Model,Schwenzer2021Review}. The advantage of MPC is that it can solve the optimization problem while respecting constraints on the MVs and MOs. It can also account for measured disturbances (MDs) and unmeasured disturbances (UDs) to the inputs and outputs. 

For this study, we use the MPC Designer tool in MATLAB \citep{MatlabVersion} for controller design. The cost function is the default used in the tool \citep{MatlabOPsite}. This function is a sum of terms scaled by weights containing the MOs, MVs, and MV increments that are tuned according to control objectives. Since one of the objectives in this study is output reference tracking, the weights corresponding to the MVs are set to \num{0}. The cost function then reduces to
\begin{equation}
\begin{split}
    J(\mathbf{z}(k)) & =\sum_{j=1}^{n_y}\sum_{i=1}^{p}\left(\tfrac{w_{i,j}^y}{s_j^y}\left[r_j(k+i|k)-y_j(k+i|k)\right]\right)^2
    \\
    & + \sum_{j=1}^{n_u}\sum_{i=0}^{p-1}\left(\tfrac{w_{i,j}^{\Delta u}}{s_j^u}\left[u_j(k+i|k)-u_j(k+i-1|k)\right]\right)^2 +\rho_{\varepsilon}\varepsilon_k^2.
\end{split}
\label{mpccostfunction}
\end{equation}
In \eqref{mpccostfunction}, $k$ is the current control interval, $p$ is the prediction horizon, $n_y$ is the number of plant outputs, and $n_u$ is the number of control inputs; here $n_u=n_y=2$. The variables $y_j(k+i|k)$, $r_j(k+i|k)$ and $u_j(k+i|k)$ are the prediction of plant output $j$, the reference value for plant output $j$, and the control input $j$, each at prediction horizon step $i$. In addition, the vector in the argument of the cost function $J$ is $\mathbf{z}(k)=[\mathbf{u}(k|k),\mathbf{u}(k+1|k),\ldots,\mathbf{u}(k+p-1|k),\varepsilon_k]^{\mathrm{T}}$, which contains the vectors $\mathbf{u}(k+i|k)= [u_1(k+i|k),\ldots,u_{n_u}(k+i|k)]^{\mathrm{T}}$ for $0\le i\le p-1$. The variable $\varepsilon_k$ is a non-negative slack variable in control interval $k$ that quantifies the constraint violation, i.e., instances in a control interval when the MOs, MVs, or MV increments cannot be held within prespecified upper and lower bounds.

The general form of the optimization problem solved by the controller at each sample time in the current control interval $k$ is
\begin{equation*}
\min_{\mathbf{z}} \ J(\mathbf{z})
\end{equation*}
such that
\begin{equation*}
\begin{array}{rccclll}
    \frac{y_{j,min}(i)}{s_j^y}-\varepsilon_k V_{j,min}^y(i) & \le
    & \frac{y_j(k+i|k)}{s_j^y} 
    & \le & \frac{y_{j,max}(i)}{s_j^y}+\varepsilon_k V_{j,max}^y(i), 
    & \quad & 
    i=1:p, \ j=1:2 \\[1em]
    \frac{u_{j,min}(i)}{s_j^u}-\varepsilon_kV_{j,min}^u(i) & \le
    & \frac{u_j(k+i-1|k)}{s_j^u} 
    & \le &\frac{u_{j,max}(i)}{s_j^u} + \varepsilon_kV_{j,max}^u(i),
    & \quad & 
    i=1:p, \ j=1:2 \\[1em]
    \frac{\Delta u_{j,min}(i)}{s_j^u}-\varepsilon_kV_{j,min}^{\Delta u}(i) & \le
    & \frac{\Delta u_j(k+i-1|k)}{s_j^u} 
    & \le &\frac{\Delta u_{j,max}(i)}{s_j^u} + \varepsilon_kV_{j,max}^{\Delta u}(i),
    & \quad & 
    i=1:p, \ j=1:2.
\end{array}
\label{mpcqpproblem}
\end{equation*}
The solution to the optimization problem is a set of $m$ control moves that minimize the cost function over the prediction horizon where $m\le p$; the integer $m$ is known as the control horizon. However, only the first control move is chosen and applied to the plant. Then, the prediction horizon is shifted to the next control interval and the procedure is repeated. If $m < p$ then some control increments are set to be zero, i.e., $u_j(k+i|k)=u_j(k+i-1|k)$ for $m+1\le i \le p$. The descriptions and values of controller tuning parameters and constraint parameters are shown in Table \ref{mpctuningparameters}. The scaling factors $s_j^y$ and $s_j^{u}$ normalize the MOs, MVs, and MV increments for improved numerical conditioning of the optimization problem.

The state-space representation of the gun turret in equations \eqref{xdotsystem} and \eqref{ysystem} is used as the plant model in the controller design. At the start of controller design, the MVs, MOs, MDs and UDs are defined in the tool. The MVs are the control inputs $u_1$ and $u_2$ and the MOs are the gun turret outputs $\theta$ and $\alpha$. Since we ignore disturbances in this study, we do not define the MDs and UDs in the design. Additionally, we leave the MVs and MV increments unconstrained, which we recognize may not be practical; however, we are not comparing controller performance to determine the `best' control approach for turret movement in this study. We are examining fundamental properties of controller errors, which are characteristics of the controlled turret system and do not depend on the input to the system.  

\begin{table} 
    \centering
    \caption{MPC tuning weights, equal constraint relaxation (ECR) parameters, MV and MO bounds, and scaling factors. Since $n_u=n_y=2$, any value reported for a parameter having a subscript $j$ indicates the value for both $j=1$ and $j=2$. Parameters having no units with their reported values are dimensionless. There are no reported values for the bounds on the MVs and MV increments because these variables are left unconstrained. }
    \begin{tabular}{cll}
        \hline
        \textbf{Parameter} & \multicolumn{1}{c}{\textbf{Description}} & \textbf{Value} \\
         \hline
         $w_{i,1}^y$ & Tuning weight for MO \num{1} at prediction horizon step $i$ & \num{1.0e2}  \\
         $w_{i,2}^y$ & Tuning weight for MO \num{2} at prediction horizon step $i$ & \num{7.5e1} \\
         $s_j^y$ & Scaling factor for MO $j$ & \SI{1.0}{[\radian]} \\
         $y_{j,min}(i)$ & Lower bound for MO $j$ at prediction horizon step $i$ & \SI{0.00}{[\radian]} \\
         $y_{1,max}(i)$ & Upper bound for MO \num{1} at prediction horizon step $i$ & \SI{6.3}{[\radian]} \\
         $y_{2,max}(i)$ & Upper bound for MO $2$ at prediction horizon step $i$ & \SI{3.1}{[\radian]} \\
         $V_{j,min}^y(i)$ & Lower ECR bound for MO $j$ at prediction horizon step $i$ & \num{1.0} \\
         $V_{j,max}^y(i)$ & Upper ECR bound for MO $j$ at prediction horizon step $i$ & \num{1.0} \\
         $w_{i,j}^{u}$ & Tuning weight for MV $j$ at prediction horizon step $i$ & \num{0.00}\\
         $w_{i,j}^{\Delta u}$ & Tuning weight for MV increment $j$ at prediction horizon step $i$ & \num{1.4e-4} \\
         $s_j^u$ & Scaling factor for MV $j$ & \SI{1.0}{[\newton\cdot\meter]} \\
         $u_{j,min}(i)$ & Lower bound for MV $j$ at prediction horizon step $i$ & -\\
         $u_{j,max}(i)$ & Upper bound for MV $j$ at prediction horizon step $i$ & -\\
         $\Delta u_{j,min}(i)$ & Lower bound for MV increment $j$ at prediction horizon step $i$ & - \\
         $\Delta u_{j,max}(i)$ & Upper bound for MV increment $j$ at prediction horizon step $i$ & - \\
         $V_{j,min}^u$ & Lower ECR bound for MV $j$ at prediction horizon step $i$ & \num{0.00} \\
         $V_{j,max}^u$ & Upper ECR bound for MV $j$ at prediction horizon step $i$ & \num{0.00}\\
         $V_{j,min}^{\Delta u}$ & Lower ECR bound for MV increment $j$ at prediction horizon step $i$ & \num{0.00} \\
        $V_{j,max}^{\Delta u}$ & Upper ECR bound for MV increment $j$ at prediction horizon step $i$ & \num{0.00} \\
         $\rho_{\varepsilon}$ & Constraint violation penalty weight & \num{1.0e5}\\
         \hline
    \end{tabular}
    \label{mpctuningparameters}
\end{table}

\subsection{Control System Objectives}
\label{controllerobjectives}
The design objectives for our controllers are to stabilize the system and rotate the gun turret to the aimpoint as quickly as possible with high accuracy. Since we consider static and moving targets in this study, we use two types of reference commands for the gun turret output to follow. For static targets, the output reference commands are step functions of the form
\begin{equation}
    r(t) = \begin{cases}
                0, &  t < 0 \\
                P, & t \ge 0
            \end{cases}
            ,
    \label{stepinput}
\end{equation}
where $P$ is either the azimuth or elevation aimpoint. For moving targets, the output reference commands are ramp functions of the form
\begin{equation}
    r(t) = \begin{cases}
                0, & t < 0  \\
                Vt, & t \ge 0
            \end{cases}
            ,
    \label{rampinput}
\end{equation}
which assumes the target is moving along a curved path at constant range with angular speed $V$.

To assess the design objective, we use the settling time $t_s$, which is a measure of the speed of the response. We choose this metric because it allows us to assess both speed and accuracy in aiming the gun turret. We define the settling time as the time it takes for the response to remain within \SI{0.10}{\percent} of the unit step function (when $R=1$ in \eqref{stepinput}). This definition is based on a level of accuracy observed in firing scenarios wherein a \SI{1}{mil} error in the initial angle of a ballistic trajectory produces a \SI{1}{\meter} miss distance at \SI{1000}{\meter} range \citep{strohm2013introduction}. Note, we use the definition of \SI{1}{mil}
defined by the North Atlantic Treaty Organization (NATO), which is an angular measurement equal to $1/6400$ of a complete circle \citep{Weaver1990system}, i.e., $1\,\si{mil}=360\si{\degree}/6400=0.05625\si{\degree}$. 
\section{Results}
\label{resultsanddiscussion}

\subsection{Experimental Setup}
\label{setup}
In this section, we present our analysis of the aiming error through multiple experiments in MATLAB \citep{MatlabVersion}. In experiment \num{1}, we examine the impact of parameter error on the aiming error distribution. In experiment \num{2}, we analyze the effects of uncertainty in the aimpoint measurement on the aiming error. In experiment \num{3}, we analyze the dependency of the mean error on the firing time. In experiment \num{4}, we model the process of measuring the aimpoint and quantify the uncertainty added to the aiming error analytically; this calculation is then compared to numerical estimates of the uncertainty from simulation data. In experiment \num{5}, we evaluate PID controller performance against moving targets.

In experiments \num{1} to \num{3}, we choose different values for the range, azimuth and elevation and run multiple sets of $10,000$ trials simulating controlled aiming at static targets. For the range, we choose nine different points uniformly from \SI{1000}{\meter} to \SI{2000}{\meter}. For the azimuth, we use six uniform values between \SI{20}{\degree} and \SI{120}{\degree}. For the elevation, we use six uniform values from \SI{5}{\degree} to \SI{30}{\degree}. These points are chosen based on typical engagement scenarios and the design limitations of modern tank systems \citep{m1abramsSpecs,lyth2021modelling}. 
For each set of trials, we choose
a firing time $t_f$, which is defined as the amount of time elapsed in a simulation of controlled gun turret movement from stationary position. Note, this is not the same as the settling time $t_s$, though the two can coincide in a simulation. The firing time is kept the same for each trial. During each trial, a target is drawn by sampling random integers between \num{1} and \num{9} for the range, \num{1} and \num{6} for the azimuth, and \num{1} and \num{6} for the elevation. The integers are sampled from uniform distributions and are used to draw target coordinates from the chosen sets of range, azimuth and elevation points. Once a target is drawn, we simulate aiming of the gun turret starting from stationary position at time $t=0$ seconds up to the firing time $t_f$; the \texttt{lsim} function is called to simulate aiming with PID control, while the \texttt{sim} function is used in simulations with MPC. In each simulation, target coordinates are used to calculate controller inputs. We set the time increment of each simulation to \num{0.01} seconds.  At the end of a set of trials, we then analyze the errors at the firing time in aggregate.

For experiments \num{1} to \num{4},
the mean of the error data is calculated as
    $\hat{\mu} = \frac{1}{N}\sum_{i=1}^N x_i,$
where $N$ is the total number of trials and $x_i$ is an observation of the error at the firing time for the trial at index $i$. The standard deviation is calculated as
    $\hat{\sigma} = \sqrt{\frac{1}{N}\sum_{i=1}^N(x_i-\hat{\mu})^2}.
    $
In all of the graphs and tables, the error data from our simulations is reported in NATO mils.

In all experiments, the PID controllers are tuned to achieve a settling time $t_s$ of \num{2} seconds according to the steps in Sections \ref{leaddesign} and \ref{pileaddesign} in the Appendix; the observed settling time is \SI{2.0}{\second} for the azimuth and elevation responses with both the lead and PI+lead controllers. This value of $t_s$ is chosen in an effort to not be too short or too long for this study. Note, after initial tuning of the PI+lead controller for the platform, we have increased the proportional gain $K_P$ by \SI{20}{\percent} to meet the settling time objective. The MPC controller is designed and tuned using the MPC Designer tool in MATLAB \citep{MatlabVersion}. We set the prediction horizon to \num{100} and the control horizon to \num{4} after experimentation. Additionally, we set a controller sample
time of \num{0.01} seconds. Since the design objective is a fast response with low error, we increase the MO weights, keep the MV weights at their default values of zero, and decrease the MV increment weights (see Table \ref{mpctuningparameters}). This tuning penalizes more for tracking error, which prioritizes output reference tracking of target coordinates rather than limiting large deviations in control inputs. The observed settling time is \SI{1.4}{\second} for the azimuth response and \SI{0.94}{\second} for the elevation response.

\subsection{Experiment 1 --- Effects of Parameter Errors on the Aiming Error}
\label{effectsParameterErrors}
In this experiment, we evaluate error distributions under more realistic operating conditions. As the plant model is only an approximation to the actual plant, there are always inherent errors in the model parameters. For that reason, this experiment assesses the effects of parameter errors (e.g., moment of inertia or damping) on the aiming error of the controlled gun turret system to accurately characterize its contribution to the error budget. 

We examine the aiming error in three targeting scenarios: assuming no error in model parameters, assuming a \SI{10}{\percent} error in the damping coefficient, and assuming a \SI{10}{\percent} error in the moment of inertia. These modifications are made separately. The error is added to a parameter as follows:
    $\hat{p} = p(1+\varepsilon)$, where $p$ is the true value of the parameter. In this experiment, we take $\varepsilon = 0.1$. Note that in each case of added error, only one model parameter is changed (damping coefficient or moment of inertia) and not the controller parameters.
    
    For the simulations, we choose \num{10} uniformly spaced firing times from \num{1} to \num{10} seconds. For each of the three models at each firing time, we conduct $10,000$ controlled aiming simulations using the PID and MPC controllers. The mean and standard deviation of the error data are calculated at the conclusion of the trials in each case. Note, since the mean of the sample could be negative, any change in the mean is discussed with respect to the absolute value. Additionally, since the mean can vary considerably depending on the samples of the reference azimuth and elevation aimpoints, we need a measure of variability that is independent of measurement units and relative to the sample mean. As a result, we calculate the coefficient of variation, which is defined as
    $
        \hat{c}_v \coloneqq \frac{\hat{\sigma}}{|\hat{\mu}|}.
    $
   
     From the results, error in the damping coefficient leads to notable changes in the aiming error distribution statistics. As seen in Table \ref{pidmeanstdevft1to6sec} for PID control, a larger mean aiming error is observed in both the azimuth and elevation beginning at a firing time of \num{3} seconds. For MPC, a larger mean aiming error in both the azimuth and elevation occurred at all firing times considered in Table \ref{mpcmeanstdevft1to6sec}. Some changes are more significant at a particular firing time depending on the firing angle considered. For instance, in the case of PID at a firing time of \num{2} seconds, Table \ref{pidmeanstdevft1to6sec} shows that the mean and standard deviation do not change for the azimuth, but do change by \SI{80}{\percent} for the elevation.
     On the other hand, for MPC at a firing time of \num{2} seconds, Table \ref{mpcmeanstdevft1to6sec} shows the mean and standard deviation change by \SI{21}{\percent} for the azimuth but only by \SI{2}{\percent} for the elevation. 
     
\begin{table}
    \centering
    \caption{Error distribution mean $\hat{\mu}$ and standard deviation $\hat{\sigma}$ for PID at different firing times. The statistics are expressed in NATO mils. The first column shows the firing time in units of seconds. For each firing time, the first row reports the results for the nominal model (N), the second row reports the results under a \SI{10}{\percent} error in the damping coefficient (DC), and the third row reports the results under a \SI{10}{\percent} error in the moment of inertia (MI).}
    \begin{tabular}{cccccc}
       \hline
     \multicolumn{2}{c}{} & \multicolumn{2}{c}{\textbf{Azimuth}} & \multicolumn{2}{c}{\textbf{Elevation}} \\
     \hline
    \textbf{Firing Time} [\si{\second}] & \textbf{Case} & $\hat{\mu}$ & $\hat{\sigma}$ & $\hat{\mu}$ & $\hat{\sigma}$ \\
       \hline
      \num{1} & \textbf{N} & \num{-.21e2} & \num{.10e2} & \num{-.65e1} & \num{.32e1} \\
     \num{1} & \textbf{DC} & \num{-.20e2} & \num{.10e2} & \num{-.70e1} & \num{.34e1} \\
      \num{1} & \textbf{MI} & \num{-.25e2} & \num{.12e2} & \num{-.67e1} & \num{.33e1} \\
    \hline
         \num{2} & \textbf{N} & \num{-.12e1} & \num{5.8e-1} & \num{-3.1e-1} & \num{1.5e-1} \\
       \num{2} & \textbf{DC} & \num{-.12e1} & \num{5.8e-1} & \num{-5.6e-1} & \num{2.7e-1} \\
       \num{2} & \textbf{MI} & \num{-0.13e1} & \num{6.1e-1} &\num{-5.5e-2} & \num{2.7e-2} \\
        \hline
         \num{3} & \textbf{N} & \num{-6.7e-2} & \num{3.3e-2} & \phantom{--} \num{1.8e-3} & \num{8.7e-4} \\
       \num{3} & \textbf{DC} & \num{-6.9e-2} & \num{3.4e-2} & \num{-6.7e-2} & \num{3.3e-2} \\
      \num{3} & \textbf{MI} & \num{-6.2e-2} & \num{3.0e-2} & \phantom{--} \num{7.2e-2}
      & \num{3.6e-2} \\
    \hline
          \num{4} & \textbf{N} & \num{-3.8e-3} & \num{1.8e-3} & \phantom{--} \num{5.7e-3} & \num{2.8e-3} \\
       \num{4} & \textbf{DC} & \num{-4.0e-3} & \num{1.9e-3} & \num{-1.3e-2} & \num{6.3e-3} \\
     \num{4} & \textbf{MI} & \num{-3.0e-3} & \num{1.5e-3} & \num{-2.5e-2} & \num{1.2e-2} \\
    \hline
          \num{5} & \textbf{N} & \num{-2.1e-4} & \num{1.0e-4} & \phantom{--} \num{1.9e-3} & \num{9.1e-4} \\
       \num{5} & \textbf{DC} & \num{-2.3e-4} & \num{1.1e-4} & \num{-3.1e-3} & \num{1.5e-3} \\
       \num{5} & \textbf{MI} & \num{-1.5e-4} & \num{7.2e-5} & \phantom{--} \num{7.1e-3} & \num{3.5e-3} \\
    \hline
         \num{6} & \textbf{N} & \num{-1.2e-5} & \num{5.8e-6} & \phantom{--} \num{5.3e-4} & \num{2.6e-4} \\
       \num{6} & \textbf{DC} & \num{-1.3e-5} & \num{6.5e-6} & \num{-8.3e-4} & \num{4.1e-4} \\
      \num{6} & \textbf{MI} & \phantom{--} \num{7.3e-6} & \num{3.6e-6} & \phantom{--} \num{2.0e-3} & \num{9.8e-4} \\
    \hline
    \end{tabular}
    \label{pidmeanstdevft1to6sec}
\end{table}
\begin{table}
    \centering
    \caption{Error distribution mean $\hat{\mu}$ and standard deviation $\hat{\sigma}$ for MPC at different firing times. The statistics are expressed in NATO mils. The first column shows the firing time in units of seconds. For each firing time, the first row reports the the results for the nominal model (N), the second row reports the results under a \SI{10}{\percent} error in the damping coefficient (DC), and the third row reports the results under a \SI{10}{\percent} error in the moment of inertia (MI).}
    \begin{tabular}{cccccc}
       \hline
     \multicolumn{2}{c}{} & \multicolumn{2}{c}{\textbf{Azimuth}} & \multicolumn{2}{c}{\textbf{Elevation}} \\
     \hline
    \textbf{Firing Time} [\si{\second}] & \textbf{Case} & $\hat{\mu}$ & $\hat{\sigma}$ & $\hat{\mu}$ & $\hat{\sigma}$ \\
       \hline
      \num{1} & \textbf{N} & \num{-.63e1} & \num{.31e1} & \phantom{--} \num{5.8e-2} & 
      \num{2.9e-2} \\
     \num{1} & \textbf{DC} & \num{-.73e1} & \num{.35e1} & \num{-1.5e-1} & \num{7.4e-2} \\
      \num{1} & \textbf{MI} & \num{-.14e2} & \num{.66e1} & \num{-7.0e-1} & \num{3.4e-1} \\
    \hline
         \num{2} & \textbf{N} & \phantom{--} \num{2.8e-2} & \num{1.4e-2} & \num{-9.6e-5} & \num{4.7e-5} \\
       \num{2} & \textbf{DC} & \phantom{--} \num{3.4e-2} & \num{1.7e-2} & \phantom{--} \num{9.8e-5} & \num{4.8e-5} \\
       \num{2} & \textbf{MI} & \phantom{--} \num{1.2e-1} & \num{5.9e-2} & \phantom{--} \num{1.4e-3} & \num{7.0e-4} \\
        \hline
          \num{3} & \textbf{N} & \num{-1.1e-4} & \num{5.1e-5} & \phantom{--} \num{7.1e-8} & \num{3.5e-8} \\
       \num{3} & \textbf{DC} & \num{-1.4e-4} & \num{6.7e-5} & \num{-6.3e-8} & \num{3.1e-8} \\
       \num{3} & \textbf{MI} & \num{-9.3e-4} & \num{4.5e-4} & \num{-2.5e-6} & \num{1.2e-6} \\
    \hline
          \num{4} & \textbf{N} & \phantom{--} \num{3.6e-7} & \num{1.7e-7} & \num{-4.1e-11} & \phantom{-}\num{2.0e-11} \\
       \num{4} & \textbf{DC} & \phantom{--} \num{4.9e-7} & \num{2.4e-7} & \phantom{--} \num{4.0e-11} & \phantom{-}\num{2.0e-11} \\
       \num{4} & \textbf{MI} & \phantom{--} \num{6.1e-6} & \num{3.0e-6} & \phantom{-}\num{4.0e-9} & \num{2.0e-9} \\
    \hline
         \num{5} & \textbf{N} & \phantom{--} \num{1.1e-9} & \phantom{-}\num{5.4e-10} & \phantom{--} 
         \num{1.7e-12} & \num{8.2e-13} \\
       \num{5} & \textbf{DC} & \phantom{--}
       \num{1.5e-9} & \phantom{-}\num{7.2e-10} & \num{-5.2e-13} & \num{1.0e-12}  \\
      \num{5} & \textbf{MI} & \num{-3.4e-8} & \num{1.7e-8} & \num{-4.4e-12} & \num{3.4e-12} \\
    \hline
        \num{6} & \textbf{N} & \phantom{-}\num{-3.8e-12} & \phantom{-}\num{7.2e-12} & \phantom{--} \num{2.0e-12} & \num{1.0e-12} \\
       \num{6} & \textbf{DC} & \phantom{---}\num{5.3e-12} & \phantom{-}\num{6.8e-12} & \num{-6.5e-13} & \num{1.6e-12} \\
     \num{6} & \textbf{MI} & \phantom{---}\num{1.3e-10} & \phantom{-}\num{6.0e-11} & \phantom{--} \num{1.7e-12} & \num{1.5e-12} \\
    \hline
    \end{tabular}
    \label{mpcmeanstdevft1to6sec}
\end{table}

As with the damping coefficient, parameter error in the moment of inertia leads to a larger mean aiming error in both firing angles, except in the case of the azimuth for PID. Despite this, error in the moment of inertia has more significant impacts on the aiming error than error in the damping coefficient. As seen in Table \ref{pidmeanstdevft1to6sec}, for PID at a firing time of \num{2} seconds, the mean of the azimuth aiming error changes by \SI{8}{\percent} and the standard deviation changes by \SI{5}{\percent}. On the other hand, the mean and standard deviation change by \SI{82}{\percent} for the elevation. The changes are more prominent for MPC at this firing time. The MPC error statistics change by a factor of \num{3.3} for the azimuth and by a factor of \num{14} for the elevation compared to PID.

The results show that parameter errors can significantly alter aiming error distribution statistics. These effects can lead to both desirable and undesirable features in the shot impact distributions by either reducing or increasing both bias and precision. Bias is how far off the mean of the impact distribution is from the desired point of impact \cite{strohm2013introduction}. Precision refers to the size of the variance of the impact distribution, with less precision indicating larger variance. Larger bias and less precision in the impact distribution can both reduce the probability of a hit.

Despite the alterations in error distribution statistics, the $c_v$ calculation in each case indicates the standard deviation remains roughly \SI{50}{\percent} of the mean for both controllers. For PID, \num{35} out of \num{36} cases have a $c_v=0.49$; only the azimuth moment of inertia case at a \num{3} second firing time  has a different value ($c_v = 0.48$). We observe a similar trend for MPC with \num{25} out of \num{36} cases having a $c_v=0.49$ and \num{4} cases differing slightly with a $c_v=0.48$; the remaining \num{7} differed significantly (e.g., the $c_v = 0.77$ for elevation moment of inertia case at a \num{5} second firing time), but these cases occur only at firing times of \num{5} and \num{6} seconds. Aside from these exceptions, which we believe are due to the controller errors reaching a numerical limit, this result is not surprising because controller errors are proportional to the reference for linear systems. From these observations we can reasonably conclude that if $\hat{\mu}_r$ and $\hat{\sigma}_r$ are the mean and standard deviation of the reference aimpoint distribution, and if $\hat{\mu}$ and $\hat{\sigma}$ are the mean and standard deviation of the error distribution, then
$\hat{\mu} = k\hat{\mu}_r$ and $\hat{\sigma}=|k|\hat{\sigma}_r$ for some $k\neq 0$. Consequently, the aiming error distribution coefficient of variation is the reference aimpoint distribution coefficient of variation. This result can be checked using the statistics of the reference aimpoint sample. Alternatively, we instead validate this result with two estimates of the proportionality constant by calculating the ratios $\hat{\mu}/\hat{\mu_r}$ and $\hat{\sigma}/\hat{\sigma_r}$ from simulation data at a firing time of  \num{2} seconds from stationary position. As seen in Table \ref{pidmpcpconstantft2sec}, the estimated proportionality constant computed from $\hat{\mu}/\hat{\mu_r}$, in absolute value, is the value of the constant calculated from $\hat{\sigma}/\hat{\sigma_r}$ in each firing scenario.

\begin{table}
    \centering
    \caption{Absolute values of the proportionality constants $\hat{\mu}/\hat{\mu}_r$ and $\hat{\sigma}/\hat{\sigma}_r$ for PID and MPC at a firing time of \num{2} seconds after starting from stationary position; the value reported in each case is the same for both constants. The first row reports the the results for the nominal model (N), the second row reports the results under a \SI{10}{\percent} error in the damping coefficient (DC), and the third row reports the results under a \SI{10}{\percent} error in the moment of inertia (MI).}
  \begin{tabular}{ccccc}
    \hline
        & \multicolumn{2}{c}{\textbf{PID}} & \multicolumn{2}{c}{\textbf{MPC}} \\
        \textbf{Case} & \textbf{Azimuth} & \textbf{Elevation} & \textbf{Azimuth} & \textbf{Elevation} 
       \\
       \hline
       \textbf{N} & \num{9.6e-4} 
       & \num{9.9e-4}
       & \num{2.2e-5}
       & \num{3.1e-7}
       \\
       \textbf{DC} & \num{9.5e-4} 
       & \num{1.8e-3}
       & \num{2.7e-5}  
       & \num{3.1e-7} 
       \\
       \textbf{MI} & \num{1.0e-3}
       & \num{1.8e-4}
       & \num{9.8e-5}
       & \num{4.6e-6}
       \\
    \hline
    \end{tabular}
    \label{pidmpcpconstantft2sec}
\end{table}

\subsection{Experiment 2 --- Effects of Aimpoint Uncertainty on the Aiming Error}
\label{effectsAimPointUncertainty}
Uncertainty in the system, e.g., in the measurement of the aimpoint or system outputs, adds additional errors that can impact the aiming error from the controlled gun turret system. To investigate the effects of uncertainty on the aiming error distributions, we consider here added measurement noise to the aimpoint, which is subsequently passed as an additional input to the control system. The noise input is considered to be random with \SI{0}{mil} mean and \SI{0.1}{mil} standard deviation. The choice for this value is based on an observed level of accuracy of \SI{1}{mil} per \SI{1}{\meter} at \SI{1000}{\meter} range \citep{strohm2013introduction}. We estimate that the level of noise from a sensor would not be \SI{1}{\percent} of \SI{1}{mil} (\SI{0.01}{mils}), nor \SI{1}{mil} either. As a result, we take the geometric mean of these two numbers which is \SI{0.1}{mils}. We then repeat the procedure in Section \ref{effectsParameterErrors}. The measurement noise is added to the input to the controlled gun turret system at the start of each simulation.

\begin{figure}
    \begin{subfigure}{0.25\textwidth}
        \centering
        \includegraphics[width=\linewidth]{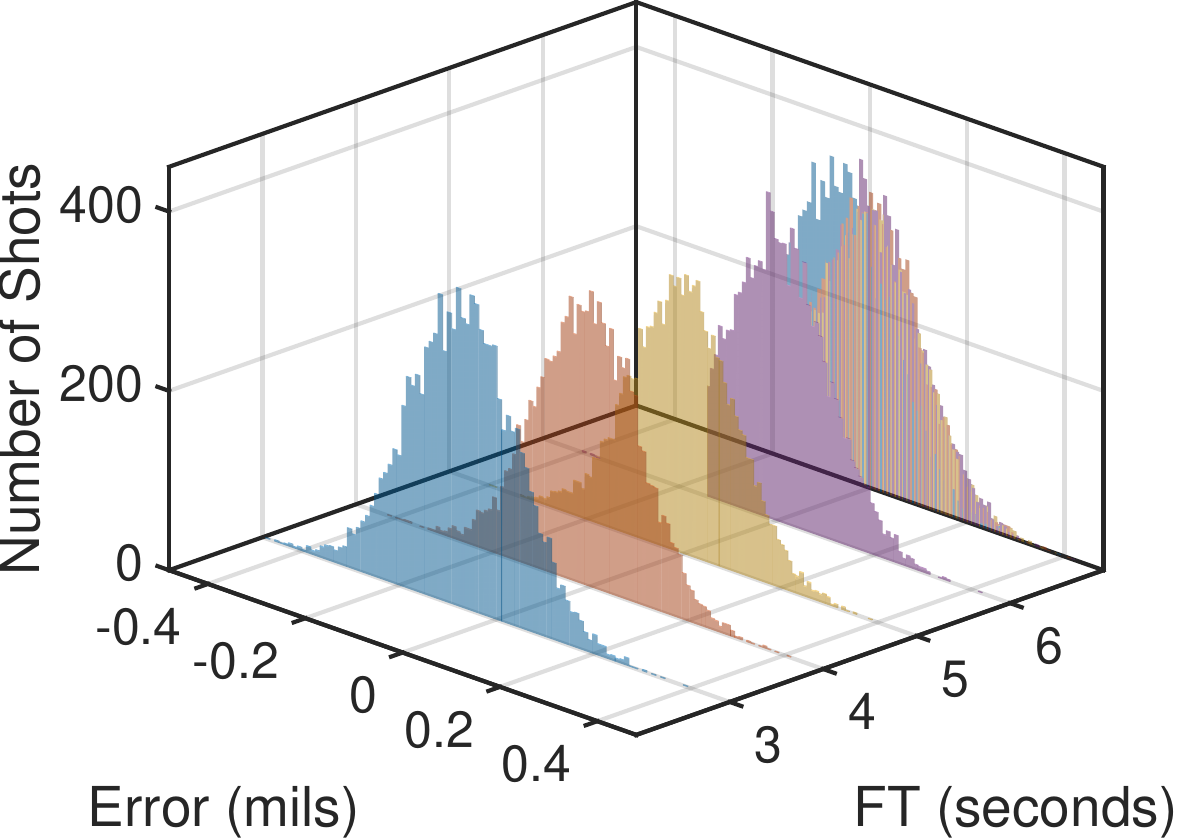}
        \caption{Azimuth}
        \label{pidazimutherrornoiseft3456}
    \end{subfigure}%
    \begin{subfigure}{0.25\textwidth}
        \centering
        \includegraphics[width=\linewidth]{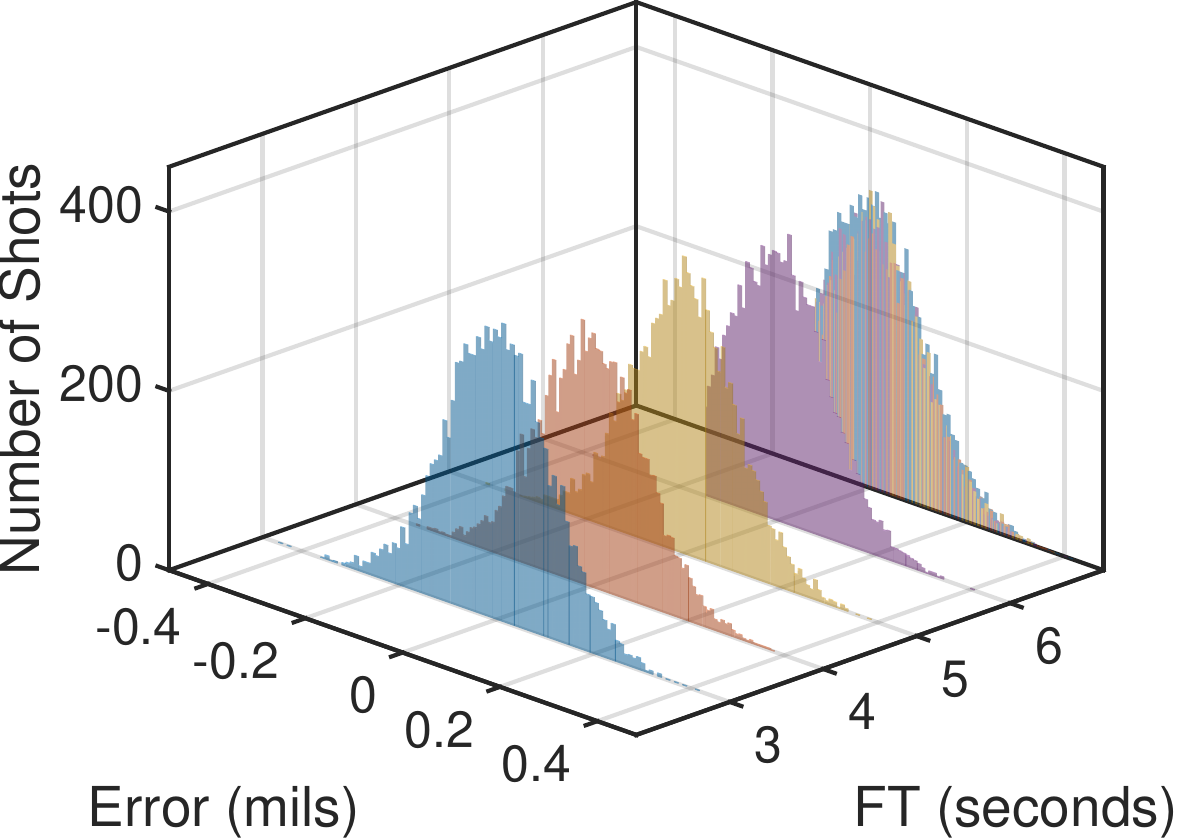}
        \caption{Elevation}
        \label{pidelevationerrornoiseft3456}
    \end{subfigure}%
    \begin{subfigure}{0.25\textwidth}
        \centering
        \includegraphics[width=\linewidth]{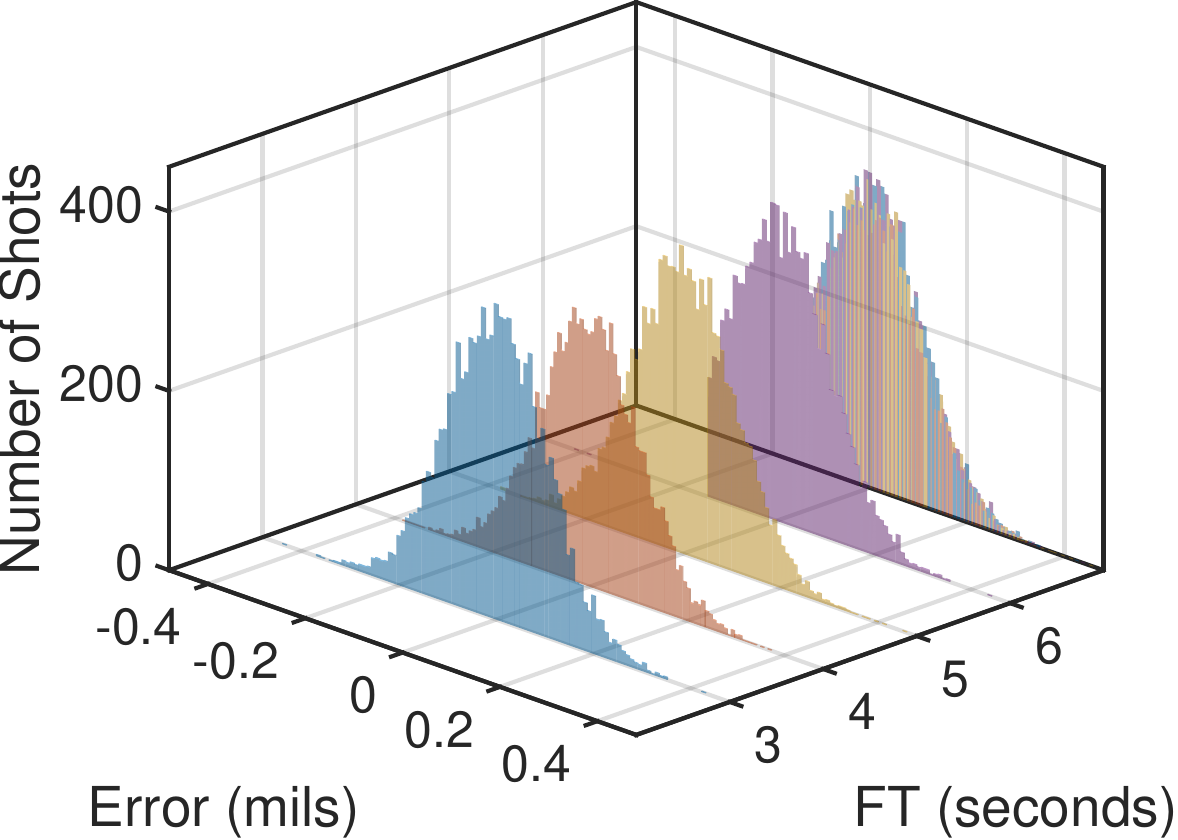}
        \caption{Azimuth}
        \label{mpcazimutherrornoiseft3456}
    \end{subfigure}%
    \begin{subfigure}{0.25\textwidth}
        \centering
        \includegraphics[width=\linewidth]{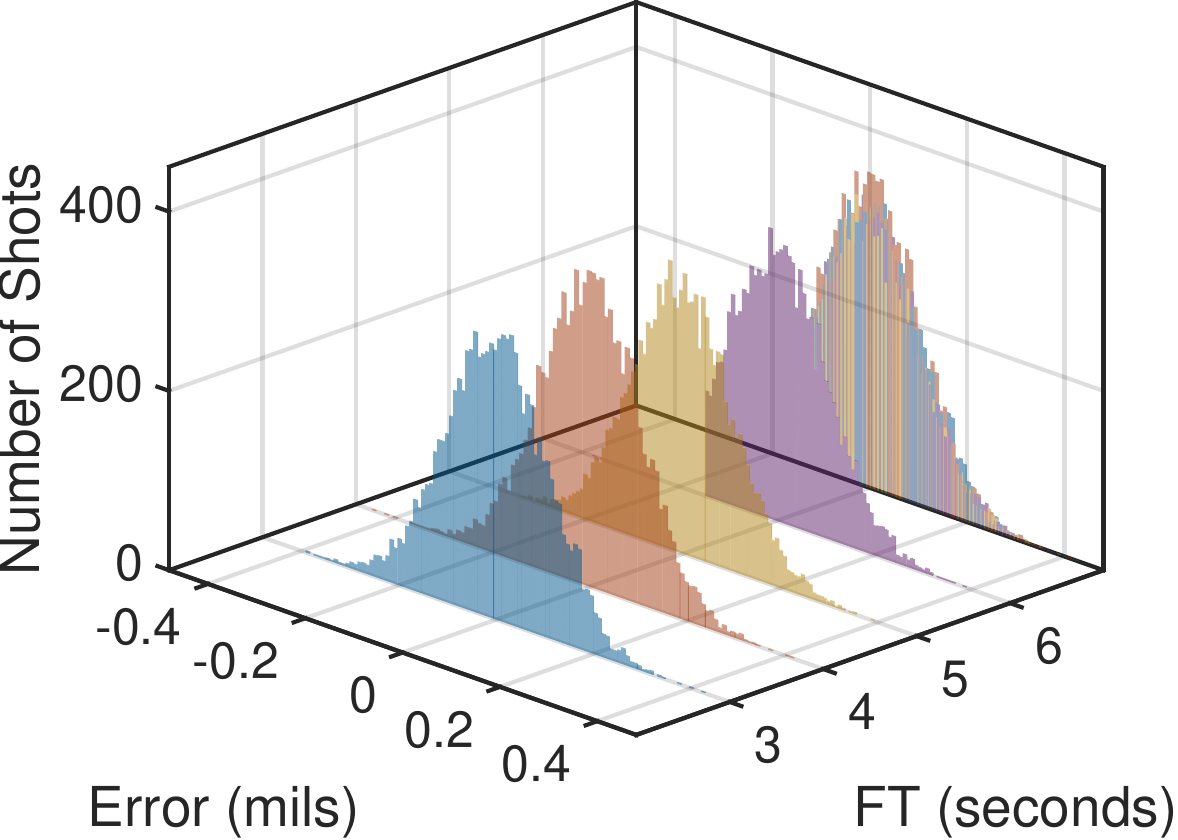}
        \caption{Elevation}
        \label{mpcelevationerrornoiseft3456}
    \end{subfigure}
    \caption{Error distributions under measurement noise with PID and MPC control for firing times of \num{3} -- \num{6} seconds. The results for PID are (\subref{pidazimutherrornoiseft3456}) and (\subref{pidelevationerrornoiseft3456}) and the results for MPC are (\subref{mpcazimutherrornoiseft3456}) and (\subref{mpcelevationerrornoiseft3456}). The noise is sampled from a normal distribution with \SI{0}{mil} mean and \SI{0.1}{mil} standard deviation. The total number of observations at each firing time is $10,000$.}
    \label{piderrornoiseft346}
\end{figure}

 In the results for PID control, it is observed that the error distributions are approaching normal distributions, as shown in Figure \ref{piderrornoiseft346}. The mean of the azimuth error distribution at a firing time of \num{3} seconds is notably shifted left from zero compared to the azimuth histograms at the three other firing times in Figure \ref{pidazimutherrornoiseft3456} and all elevation histograms in Figure \ref{pidelevationerrornoiseft3456}. This can also be seen in Table \ref{pidmeannoise} as the mean at this firing time is roughly \num{10} to \num{100} times larger in absolute value than the mean azimuth error after a \num{3} second firing time and the mean elevation error after a \num{2} second firing time. One reason for this difference is that the elevation errors are contained in a smaller range of values near the settling time of the controller since target elevation coordinates are limited to \SI{30}{\degree} and the standard deviation of the noise input is the same for both the azimuth and elevation. After \num{3} seconds, however, there is not much change between the distributions obtained from firing at \num{4} seconds, \num{5} seconds, and \num{6} seconds from stationary position. Likewise, there is not much change between all three elevation error distributions in Figure \ref{pidelevationerrornoiseft3456} either. Additionally, Table \ref{pidmeannoise} shows that the mean of both the azimuth and elevation error distributions are approaching the mean of the noise input, which is expected since controller errors have zero mean at steady-state \citep{Kluever2015Ch8Fvt}. We also observe this trend in the standard deviation of the azimuth and elevation error distributions for firing times after \num{2} seconds from stationary position; for \num{6} out of the \num{8} cases from firing times of \num{3} seconds to \num{6} seconds they are \SI{0.11}{mils}, which is within \SI{10}{\percent} of the noise input standard deviation of \SI{0.1}{mils}, and for the remaining 2 cases they are \SI{0.1}{mils}.
 
 For MPC, as can be seen in Figure \ref{mpcazimutherrornoiseft3456} and Figure \ref{mpcelevationerrornoiseft3456}, the azimuth and elevation error distributions are almost the same at all four firing times; they are also approaching normal distributions, as with PID control. In addition, the mean azimuth and elevation errors are closer to zero than the mean errors for PID control at the \num{3} second firing time, which is confirmed in Table \ref{mpcmeannoise}. But this is due to the azimuth and elevation responses each having a faster settling time with MPC. We similarly observe in Table \ref{mpcmeannoise} that the mean azimuth and elevation errors are approaching the mean of the noise input, as with PID control. Likewise, we observe this trend with the standard deviations of the azimuth and elevation error distributions after a \num{2} second firing time from stationary position; they are all \SI{0.10}{mils}, which is the standard deviation of the noise input. 
 
 The results for both PID and MPC control indicate that measurement noise can alter both the mean and standard deviation of the error distribution. This can lead to a reduction in firing accuracy depending on the level of noise and firing time.

\begin{table}
    \centering
    \caption{Error distribution mean and noise input mean for PID at six different firing times. The statistics are expressed in NATO mils. The first the column shows the firing time in units of seconds. The noise is sampled from a normal distribution with \SI{0}{mil} mean and \SI{0.1}{mil} standard deviation. The total number of observations at each firing time is $10,000$.}
    \begin{tabular}{ccccc}
       \hline
        & \multicolumn{2}{c}{\textbf{Azimuth}} & \multicolumn{2}{c}{\textbf{Elevation}} \\
        \hline
        \textbf{Firing Time} [\si{\second}] & \textbf{Output} & \textbf{Noise Input} & \textbf{Output} & \textbf{Noise Input} \\
       \hline
       \num{1} & \num{-0.21e2} & \num{-1.4e-4} & \num{-0.65e1} & \phantom{--} \num{1.7e-3} \\
       \num{2} & \num{-0.12e1} & \phantom{--} \num{8.6e-4} & \num{-3.1e-1} & \num{-1.6e-3} \\ 
       \num{3} & \num{-6.8e-2} & \num{-8.1e-4} & \phantom{--} \num{3.0e-3} & \phantom{--} \num{8.2e-4} \\
       \num{4} & \num{-4.7e-3} & \num{-6.1e-4} & \phantom{--} \num{5.8e-3} & \phantom{--} \num{4.5e-4} \\
       \num{5} & \num{-7.4e-4} & \num{-4.5e-4} & \phantom{--} \num{1.5e-3} & \num{-5.8e-4} \\
       \num{6} & \num{-6.6e-4} & \num{-7.3e-4} & \num{-1.7e-3} & \num{-2.5e-3} \\
    \hline
    \end{tabular}
    \label{pidmeannoise}
\end{table}
\begin{table}
    \centering
    \caption{Error distribution mean and noise input mean for MPC at six different firing times. The statistics are expressed in NATO mils. The first column shows the firing time in units of seconds. The noise is sampled from a normal distribution with \SI{0}{mil} mean and \SI{0.1}{mil} standard deviation. The total number of observations at each firing time is $10,000$.}
    \begin{tabular}{ccccc}
       \hline
        & \multicolumn{2}{c}{\textbf{Azimuth}} & \multicolumn{2}{c}{\textbf{Elevation}} \\
        \hline
       \textbf{Firing Time} [\si{\second}] & \textbf{Output} & \textbf{Noise Input} & \textbf{Output} & \textbf{Noise Input} \\
        \hline
       \num{1} & \num{-0.63e1} & \num{-1.9e-3} & \phantom{--} \num{6.1e-2} & \phantom{--} \num{2.9e-3} \\
       \num{2} & \phantom{--} \num{2.8e-2} & \phantom{--} \num{2.1e-4}& \phantom{--} \num{7.2e-5} & \phantom{--} \num{3.1e-4} \\ 
       \num{3} & \num{-1.3e-4} & \phantom{--} \num{2.1e-4} & \num{-1.0e-3} & \num{-9.8e-4} \\
       \num{4} & \num{-1.1e-3} & \num{-1.2e-3} & \phantom{--} \num{6.1e-5} & \num{-3.0e-5} \\
       \num{5} & \phantom{--} \num{1.1e-3} & \phantom{--} \num{7.9e-4} & \num{-1.9e-3} & \num{-1.7e-3} \\
       \num{6} & \phantom{--} \num{1.6e-3} & \phantom{--} \num{1.5e-3} & \num{-9.6e-4} & \num{-9.4e-4} \\
    \hline
    \end{tabular}
    \label{mpcmeannoise}
\end{table}

\subsection{Experiment 3 --- Mean Aiming Error vs Firing Time}
\label{meanerrvsft}

The firing time impacts the aiming error. Knowing the dependency of the mean and standard deviation of the error distributions on the firing time can inform decisions on when to engage a target and also controller design. To quantify this dependency, we examine the mean and standard deviation of the error data acquired from the controlled gun turret simulations in experiments \num{1} and \num{2} for four cases: no error in model parameters or measurement noise, \SI{10}{\percent} error in the damping coefficient, \SI{10}{\percent} error in the moment of inertia, and measurement noise. 

\begin{figure}
    \begin{subfigure}{0.25\textwidth}
        \centering
        \includegraphics[width=\linewidth]{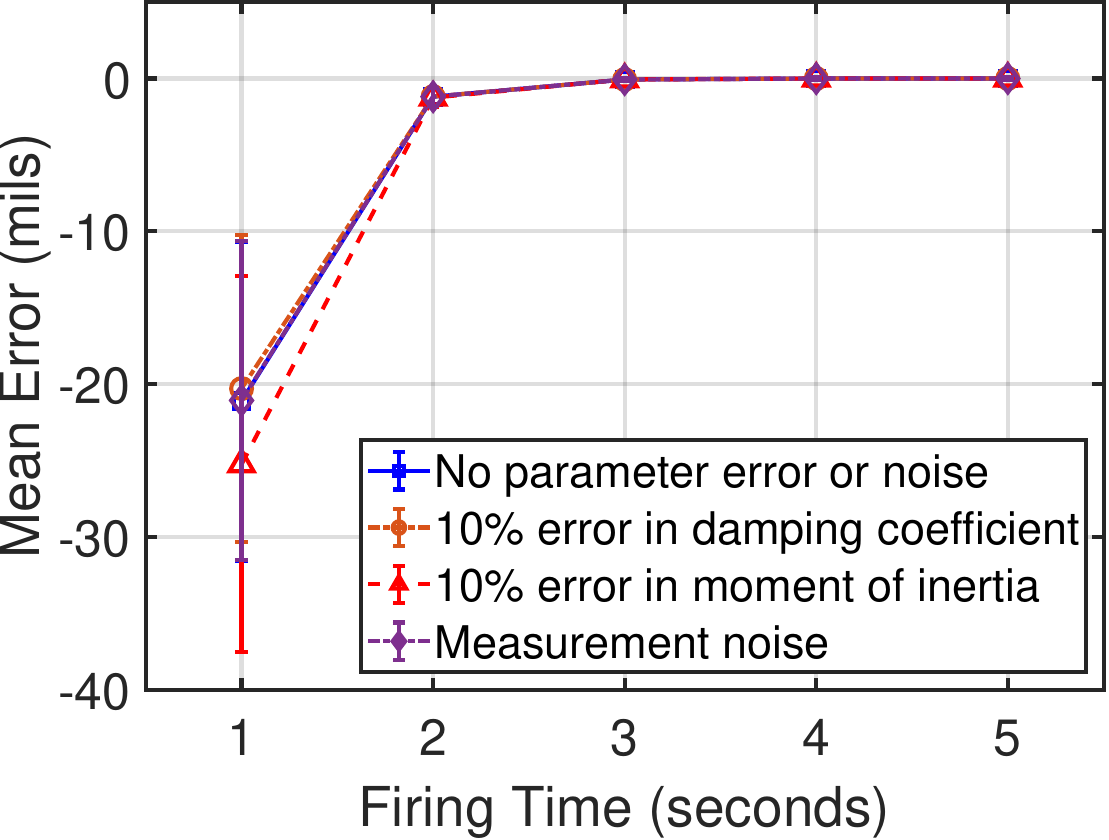}
        \caption{Azimuth}
        \label{meanerrazimuthvsftpid}
    \end{subfigure}%
    \begin{subfigure}{0.25\textwidth}
        \centering
        \includegraphics[width=\linewidth]{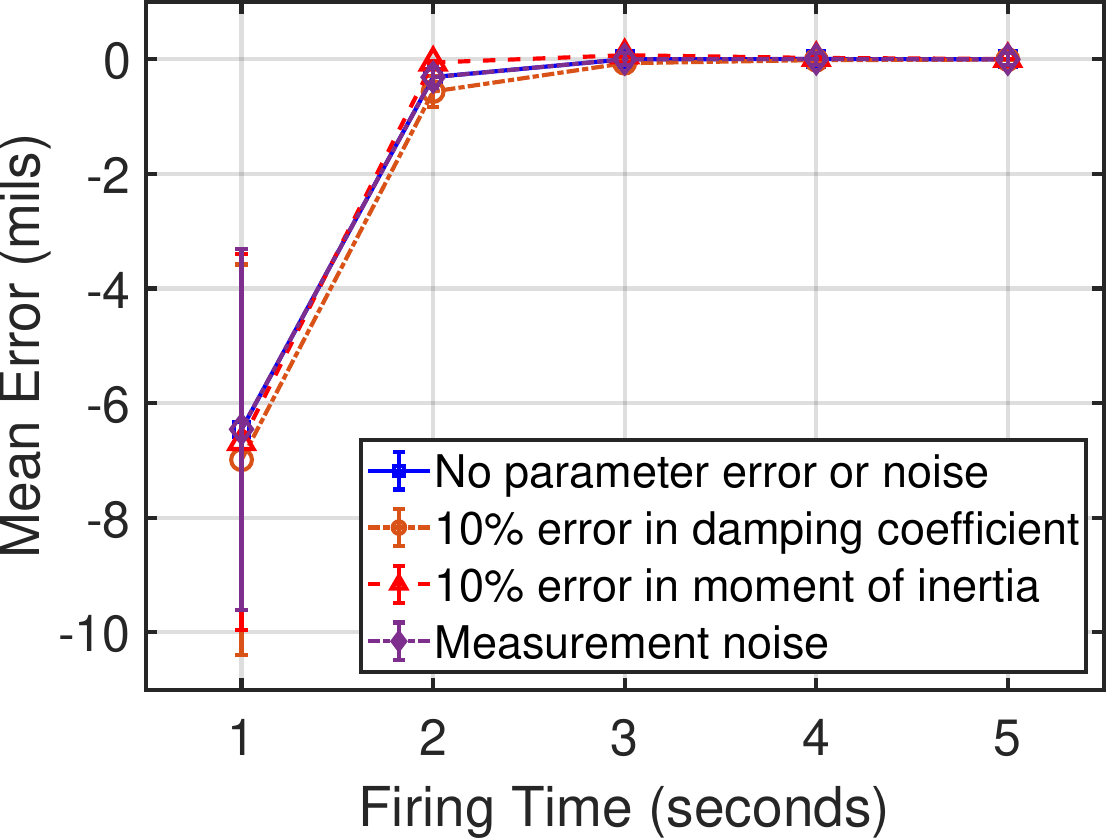}
        \caption{Elevation}
        \label{meanerrelevationvsftpid}
    \end{subfigure}%
    \begin{subfigure}{0.25\textwidth}
        \centering
        \includegraphics[width=\linewidth]{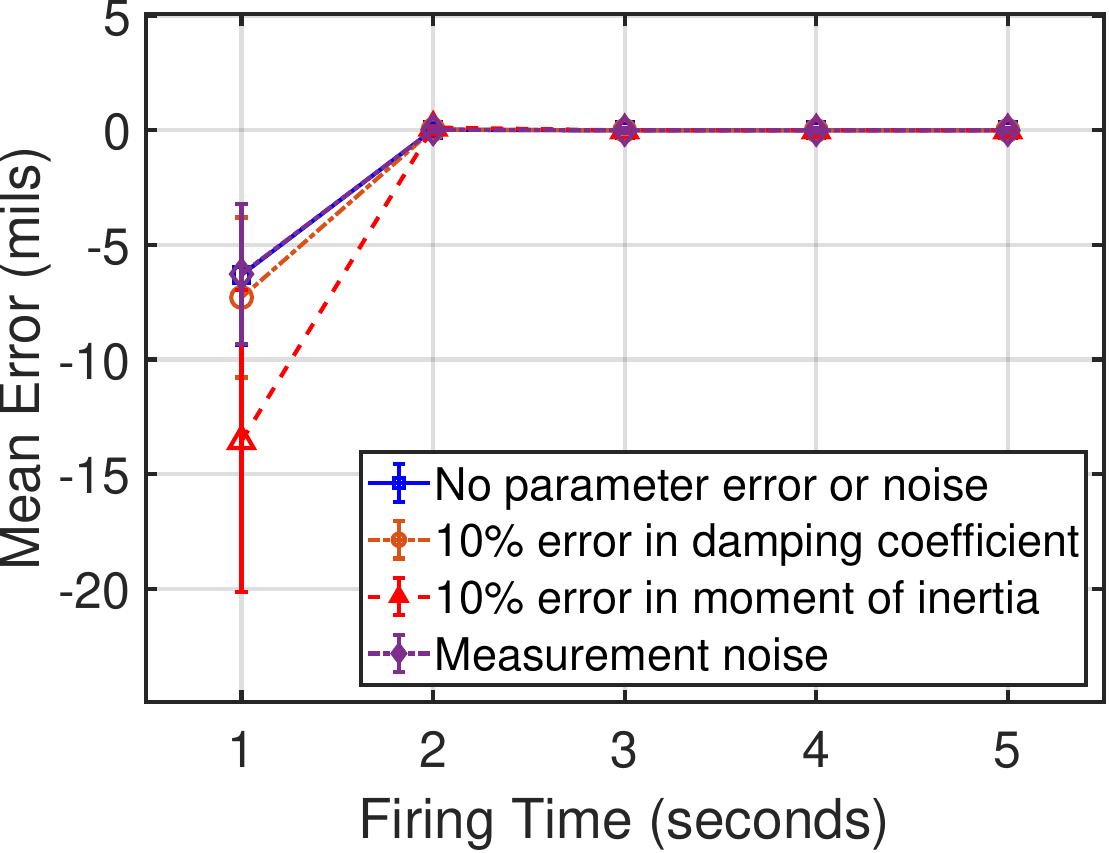}
        \caption{Azimuth}
        \label{meanerrazimuthvsftmpc}
    \end{subfigure}%
    \begin{subfigure}{0.25\textwidth}
        \centering
        \includegraphics[width=\linewidth]{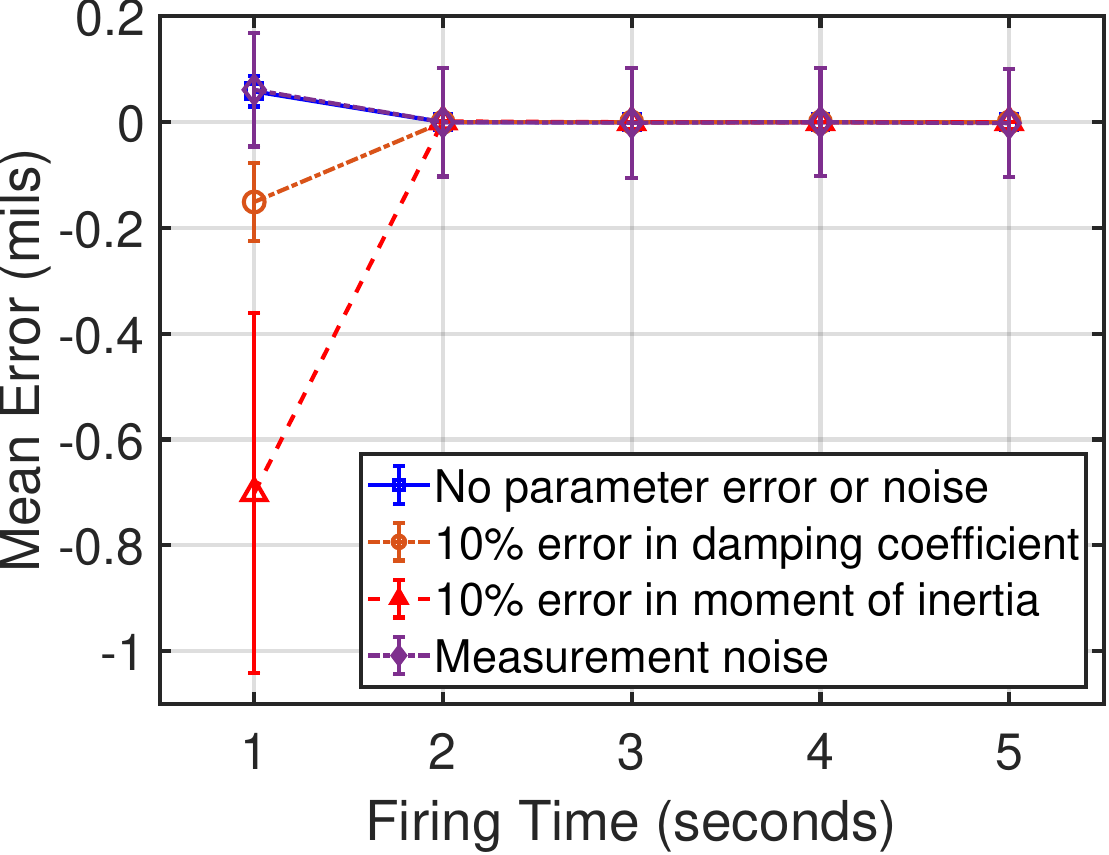}
        \caption{Elevation}
        \label{meanerrelevationvsftmpc}
    \end{subfigure}
    \caption{Mean error as function of firing time in four different targeting scenarios for PID and MPC control. The results for PID are (\subref{meanerrazimuthvsftpid}) and (\subref{meanerrelevationvsftpid}) and the results MPC are (\subref{meanerrazimuthvsftmpc}) and (\subref{meanerrelevationvsftmpc}). In the case of measurement noise, the noise is sampled from a normal distribution with \SI{0}{mil} mean and \SI{0.1}{mil} standard deviation.}
    \label{meanerrvsftpidmpc}
\end{figure}

The results for PID control show that the mean errors are approaching zero in three of the cases for the azimuth and elevation in Figure \ref{meanerrazimuthvsftpid} and Figure \ref{meanerrelevationvsftpid}: no error in model parameters or measurement noise (blue curve), \SI{10}{\percent} error in the damping coefficient (orange curve), and \SI{10}{\percent} error in the moment of inertia (red curve); this is confirmed in Table \ref{pidmeanstdevft1to6sec}. As for the case of measurement noise (purple curve), the mean error is approaching the mean of the noise input, which is confirmed in Table \ref{pidmeannoise}. This is because the number of noise samples remains the same at each firing time. As with the mean, the standard deviations, indicated by the size of the error bars, are approaching zero in the same three cases of parameter error for the azimuth and elevation in Figure \ref{meanerrazimuthvsftpid} and Figure \ref{meanerrelevationvsftpid}, which is also corroborated in Table \ref{pidmeanstdevft1to6sec}. In the case of measurement noise (purple curve), the standard deviations are approaching that of the noise input for the azimuth and elevation, in agreement with Table \ref{pidmeannoise}.

The mean error for MPC control in Figure \ref{meanerrvsftpidmpc} shows similar trends as in the case of PID control. The difference is that the mean error in the three cases of parameter error at each firing time is smaller in absolute value than the corresponding mean error for PID control, which is verified in Table \ref{mpcmeanstdevft1to6sec}; moreover, the mean errors in these cases are approaching zero more rapidly than for PID control. These observations are due to assuming an ideal plant model in the controller design and tuning the controller for a faster settling time. In the case of measurement noise, once again, we observe that the mean error and standard deviation are approaching the mean and standard deviation of the noise input, which agrees with the results in Table \ref{mpcmeannoise}.

The results in Table \ref{pidmeanstdevft1to6sec}, Table \ref{mpcmeanstdevft1to6sec}, and Figure \ref{meanerrvsftpidmpc} show that in the absence of measurement noise, the mean and standard deviation of the error distributions approach \num{0} the longer one waits to fire. However, this result is under the assumption of a linear system, static targets, and no time constraints for rotating the gun turret to the aimpoint. In this case, controller errors are proportional to target coordinates at a given firing time, which is in accordance with the proportionality constants at a firing time of \num{2} seconds in Table \ref{pidmpcpconstantft2sec}. It follows that this result is consistent with control theory because the theoretical steady-state errors are \num{0} for the gun turret under lead control or PI+lead control \citep{Kluever2015Ch8Fvt}. It is also evident that firing before the settling time of the controller can lead to larger errors. For both PID and MPC, the standard deviations are larger for faster firing times since the transient part of the response has not had enough time to significantly decay. As with the results in Section \ref{effectsAimPointUncertainty}, this could lead to a reduction in firing accuracy depending on when one chooses to fire.

\subsection{Experiment 4 --- Model of Aimpoint Uncertainty and the Aiming Error }
\label{modelAimPointUncertainty}
This experiment further explores the effects of uncertainty in the aimpoint measurement on the aiming error by comparing analytical calculations of error distribution statistics with numerical estimates of the statistics from simulation data. However, instead of simply adding noise to the measured aimpoint, as we have done in experiment 2, we model the effect of the noise as a random process with dynamics similar to an AI target recognition system. At time $t\ge 0$, we assume the reference input $r(t)$ is the response to $w(t)$, which is a random variable sampled from a normal distribution with \SI{0}{mil} mean and standard deviation $\sigma_w$. The transfer function model between $w(t)$ and $r(t)$ in the Laplace domain is chosen as the following: 
\begin{equation}
    R(s) = \frac{1}{\tau s + 1}W(s),
    \label{whitenoise2ref}
\end{equation}
where $R(s)$ is the Laplace transform of $r(t)$ and $W(s)$ is the Laplace transform of $w(t)$; $s$ is the Laplace variable in units of \si{\radian/\second}. The system time constant is $\tau$, which is the speed of the dynamics. The transfer function in \eqref{whitenoise2ref} can be interpreted as a model of the effects of noise imparted by an AI target recognition system as it updates target location in real-time, and the time constant, an estimate of the speed the system can analyze images to extract target coordinates. We assume that this procedure occurs at \SI{0.5}{\hertz}, and accordingly, we set the time constant to $\tau = 2$ seconds. 

The transfer function in \eqref{whitenoise2ref} is used to determine the relationship between $w(t)$ and the error $e(t)$ in rotating the controlled gun turret to the aimpoint. Using a block diagram of the closed-loop system, the error in the Laplace domain is
\begin{equation}
    E(s) = \frac{1}{1+K(s)G(s)}R(s),
    \label{errorsystemtf}
\end{equation}
where $K(s)$ and $G(s)$ are the controller and gun turret system transfer functions, respectively. By substituting \eqref{whitenoise2ref} into \eqref{errorsystemtf}, the transfer function between
$w(t)$ and $e(t)$ is the rational function multiplying $W(s)$ in the following:
\begin{equation}
    E(s) = \frac{1}{(2s+1)(1+K(s)G(s))}W(s).
    \label{w2errorsystem}
\end{equation}
Note there are two transfer functions of the form of the rational function in \eqref{w2errorsystem} for the azimuth and elevation subsystems. 

Since the error systems are linear and the random variable $w(t)$ is normally distributed, the error responses to $w(t)$ are also normally distributed. As a result, the theoretical probability density function of the error $e(t)$ is the normal distribution as shown in the following equation:
\begin{equation}
    \rho(e) = \frac{1}{\sigma_e\sqrt{2\pi}}\exp{\left(-\tfrac{1}{2}\tfrac{(e-\mu_{e})^2}{\sigma_{e}^2}\right)},
    \label{gaussianpdfe}
\end{equation}
where $\mu_e$ and $\sigma_e$ are the mean and standard deviation of the distribution. 

From the Final Value Theorem \citep{Kluever2015Ch8Fvt}, the steady-state errors against static targets are zero for both the azimuth and elevation closed-loop systems. Thus, the mean of their theoretical error distributions at steady-state is $\mu_e = $ \SI{0}{mils}. 

Using results from linear system theory, the standard deviation of the distributions can be determined from the 2-norm of the error system transfer function in \eqref{w2errorsystem}. In discrete time, the error system 2-norm is defined as
\begin{equation}
    \Vert H\Vert_2 \coloneqq \sqrt{\frac{1}{2\pi}\int_{-\omega_N}^{\omega_N}|H(e^{j\omega h)}|^2\,d\omega},
    \label{2norm}
\end{equation}
where
$H$ is the transfer function in \eqref{w2errorsystem}, $\omega_N$ is the Nyquist frequency, or half the sampling frequency in \si{\radian/\second}, and $h$ is the sample time in seconds. Given the white noise input $w(t)$, the standard deviation of the error distributions at steady-state is
\begin{equation}
    \sigma_{e} = \Vert H\Vert_{2}\frac{\sigma_{w}}{\sqrt{f_s}},
    \label{stdeverror}
\end{equation}
where $f_s=1/h$ is the sampling frequency. See Section \ref{errorstdevderivation} in the Appendix for a derivation of this equation.

In the numerical experiment, the firing time is chosen at the start and applied in each iteration to measure the average error. The mean and standard deviation of the input data and error data are calculated as described in Section \ref{setup}. We then compare the error distribution mean and standard deviation calculated from simulation data with the theoretical mean $\mu_e = $ \SI{0}{mils} and standard deviation $\sigma_e$ given in \eqref{stdeverror}. 

For the comparison, the simulations are performed with the PID controllers. We run $10,000$ trials simulating the error response to the white noise input $w(t)$ using the transfer function in \eqref{w2errorsystem} converted to discrete time. The sampling period is chosen to be $h = 0.02$ seconds. We assume a \SI{1}{mil} standard deviation for the white noise. The firing time is set to \num{10} seconds after stationary position so that the system output is in steady-state. At the conclusion of the trials, we then calculate the mean and standard deviation of the input data and error data at the firing time over all trials. The numerical mean and standard deviation of the error distributions are compared with their theoretical counterparts calculated with equations \eqref{2norm} and \eqref{stdeverror}. We also compare a numerical calculation of the error system 2-norms with the theoretical result in \eqref{2norm}. The 2-norm is calculated numerically as in the following:
\begin{equation}
    \Vert \hat{H}\Vert_2 = \frac{\hat{\sigma}_e}{\hat{\sigma}_w}\sqrt{f_s}.
\end{equation}

\begin{figure}
    \begin{subfigure}{0.5\textwidth}
        \centering
        \includegraphics[width=0.7\linewidth]{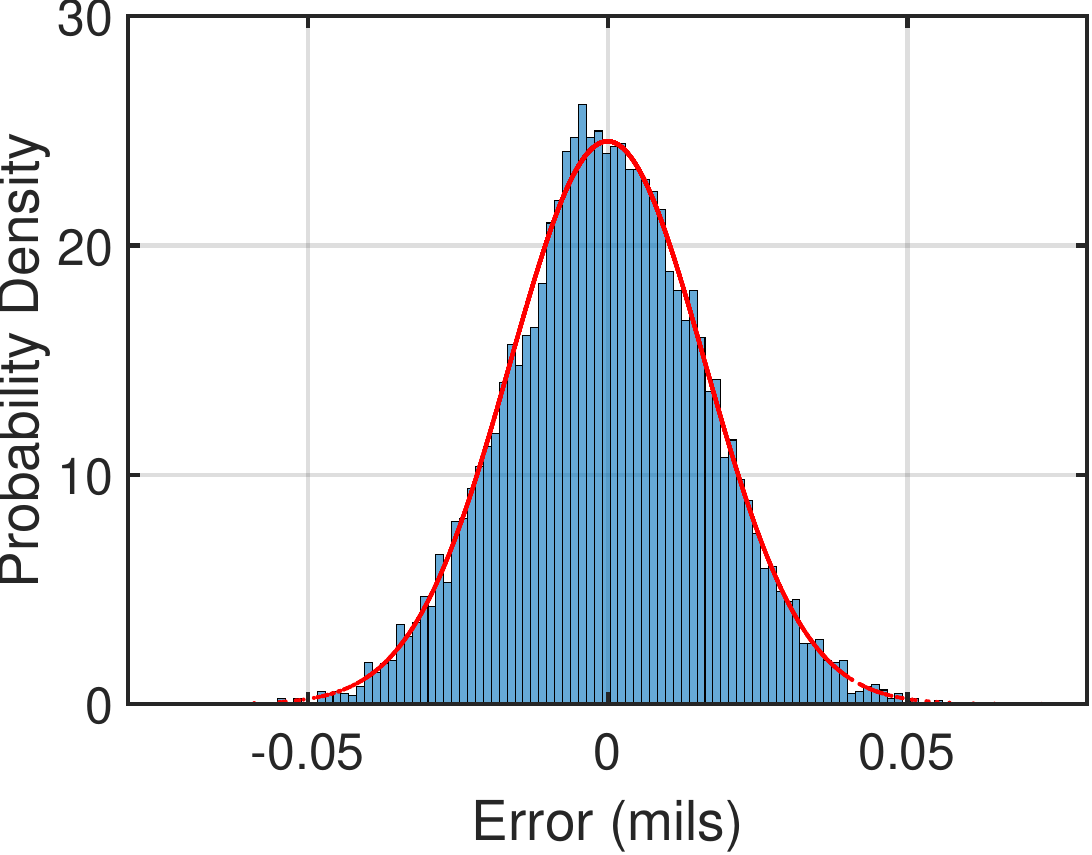}
        \caption{Azimuth}
        \label{pidazimuthpdfnoiseft10}
    \end{subfigure}%
    \begin{subfigure}{0.5\textwidth}
        \centering
        \includegraphics[width=0.7\linewidth]{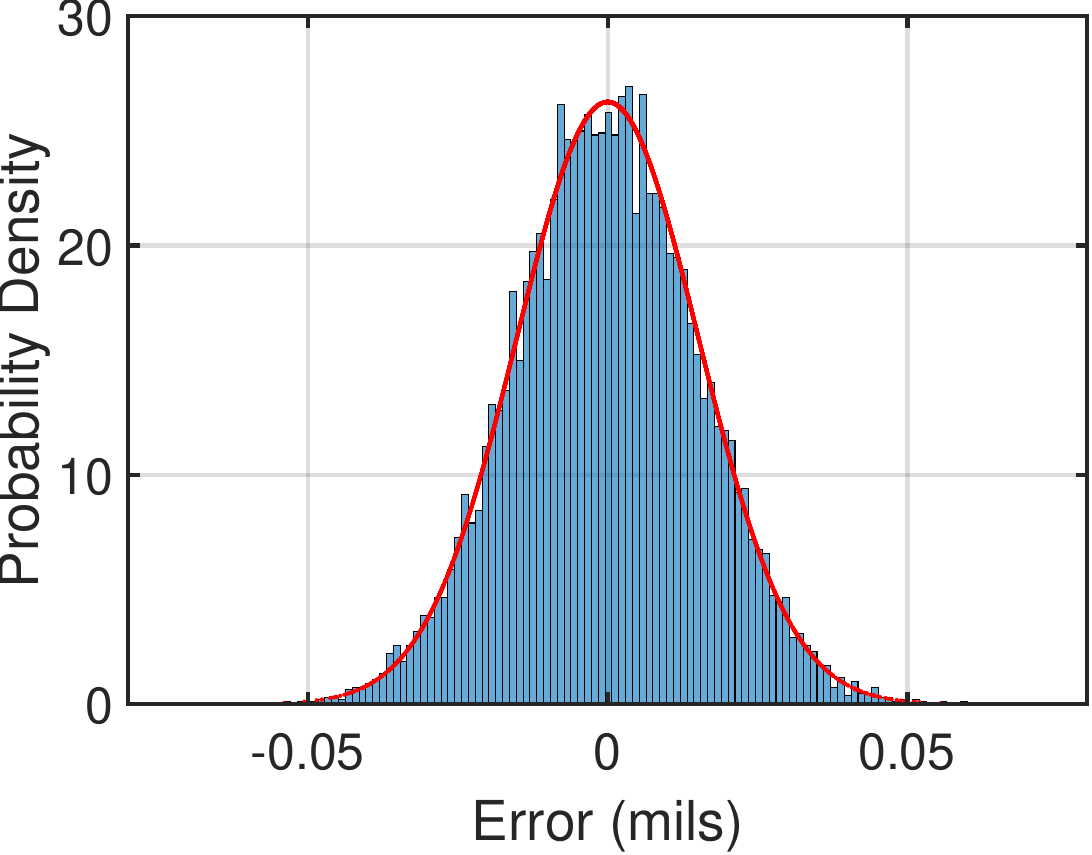}
        \caption{Elevation}
       \label{pidelevationpdfnoiseft10}
    \end{subfigure}
    \caption{Error distributions for PID control normalized for probability density. The firing time is \num{10} seconds from stationary position. Target measurement noise is sampled from a normal distribution with \SI{0}{mil} mean and \SI{1}{mil} standard deviation. The red scatter plot is the theoretical probability density function obtained by evaluating \eqref{gaussianpdfe} with the error data. The total number of observations is $10,000$.}
    \label{pidpdfnoiseft10}
\end{figure}

The results show good agreement between the numerical and theoretical calculations of the error distribution statistics and system 2-norms. In Table \ref{piderrorstats}, the numerical means agree with the theoretical values up to three decimal places for both the azimuth and elevation. The numerical standard deviations agree with the theoretical values to within \SI{0.0011}{\percent} for the azimuth and \SI{0.8}{\percent} for the elevation. The numerical error system 2 norms agree to within \SI{0.3}{\percent} for the azimuth and \SI{1.1}{\percent} for the elevation. The error distributions at a firing time of \num{10} seconds plotted from the data, normalized for probability density, appear to follow a plot of the theoretical result \eqref{gaussianpdfe} (red scatter plot) in Figure \ref{pidazimuthpdfnoiseft10} and Figure \ref{pidelevationpdfnoiseft10}. 

\begin{table}
    \centering
    \caption{Error distribution mean and standard deviation and error system \num{2}-norms for PID. The firing time is \num{10} seconds from stationary position. The statistics are expressed in NATO mils. Quantities marked with a carrot are calculated from simulation data and those without are calculated analytically. Note the theoretical means are \num{0} while the simulation generated very small deviations from \num{0}.}
    \begin{tabular}{ccccccc}
        \hline
    & $\hat{\mu}$ &  $\mu$ & $\hat{\sigma}$ & $\sigma$ & $\Vert \hat{H}\Vert_2$ & $\Vert H\Vert_2$ \\
        \hline
       \textbf{Azimuth} & \phantom{$-$}\num{5.762e-5} & \num{0.0000} & \num{0.01625} & \num{0.01625} & \num{0.1152} & \num{0.1149} \\
       \textbf{Elevation} & \num{-1.055e-4} & \num{0.0000} & \num{0.01531} & \num{0.01519} & \num{0.1086} & \num{0.1074} \\
       \hline
    \end{tabular}
    \label{piderrorstats}
\end{table}

\subsection{Experiment 5 --- Moving Targets}
\label{movingtargets}
The results of the previous sections have revealed important aspects of the aiming error \linebreak distributions from controlled weapon systems. So far we considered static targets. In this experiment, 
we consider a moving target scenario and evaluate the controller performance. 

For the experiment, we model a moving target as a ramp input, similar to \citet{Ma2022adaptive}.  
We assume that the target is moving at an angular speed of \SI{10}{\degree/\second} in the directions of increasing azimuth and increasing elevation. We first evaluate the performance of the PID controllers used in experiments \num{1} to \num{4} for static targets, which are a lead controller for the azimuth and a PI+lead controller for the elevation. Recall that these controllers are designed to meet a settling time of \num{2} seconds according to the procedures in Sections \ref{leaddesign} and \ref{pileaddesign} of the Appendix. We then design a PI+lead controller for the azimuth and evaluate the performance. This controller is also designed to meet a settling time of 2 seconds.
\begin{figure}[!hb]
	\begin{subfigure}{0.32\textwidth}
		\centering
		\includegraphics[width=0.9\linewidth]{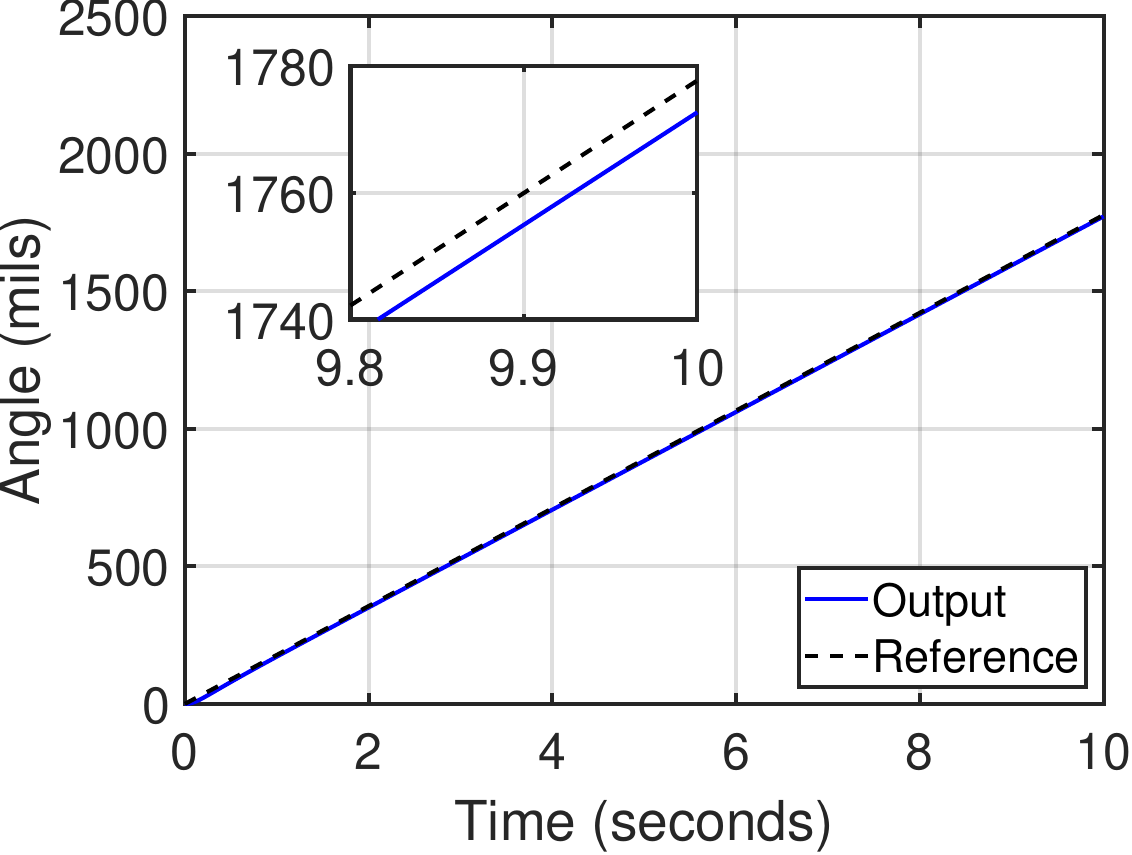}
		\caption{Azimuth}
		\label{pidazimuthleadmtoutput}
	\end{subfigure}
	\begin{subfigure}{0.32\textwidth}
		\centering
		\includegraphics[width=0.9\linewidth]{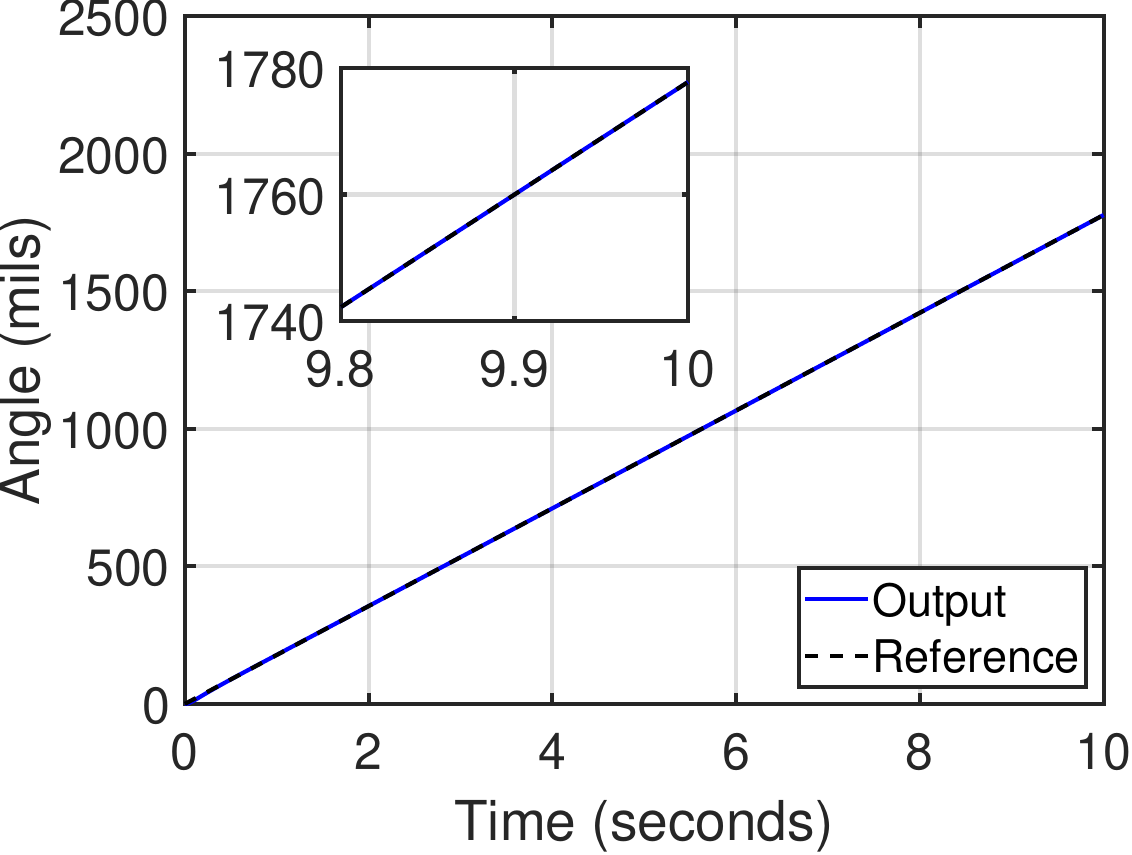}
		\caption{Azimuth}
		\label{pidazimuthpileadmtoutput}
	\end{subfigure}
	\begin{subfigure}{0.32\textwidth}
		\centering
		\includegraphics[width=0.9\linewidth]{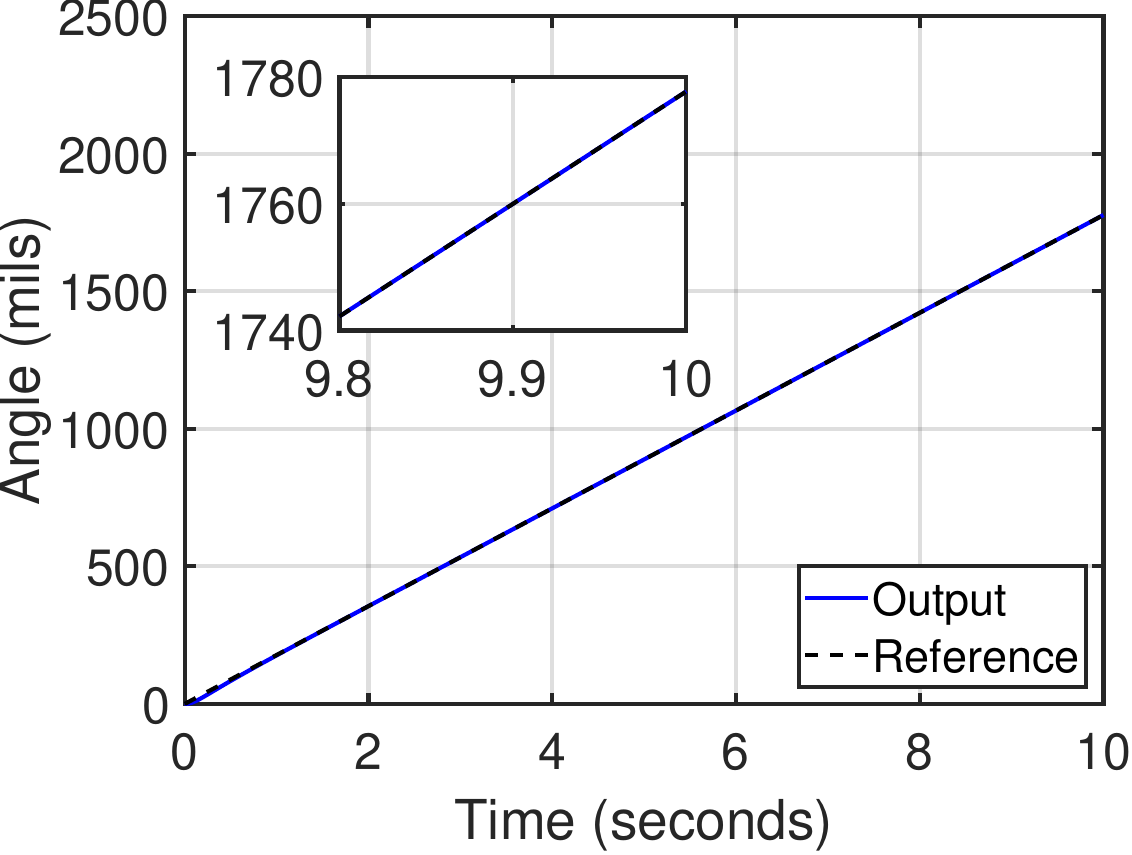}
		\caption{Elevation}
		\label{pidelevationpileadmtoutput}
	\end{subfigure}
	\\
	\begin{subfigure}{0.32\textwidth}
		\centering
		\includegraphics[width=0.9\linewidth]{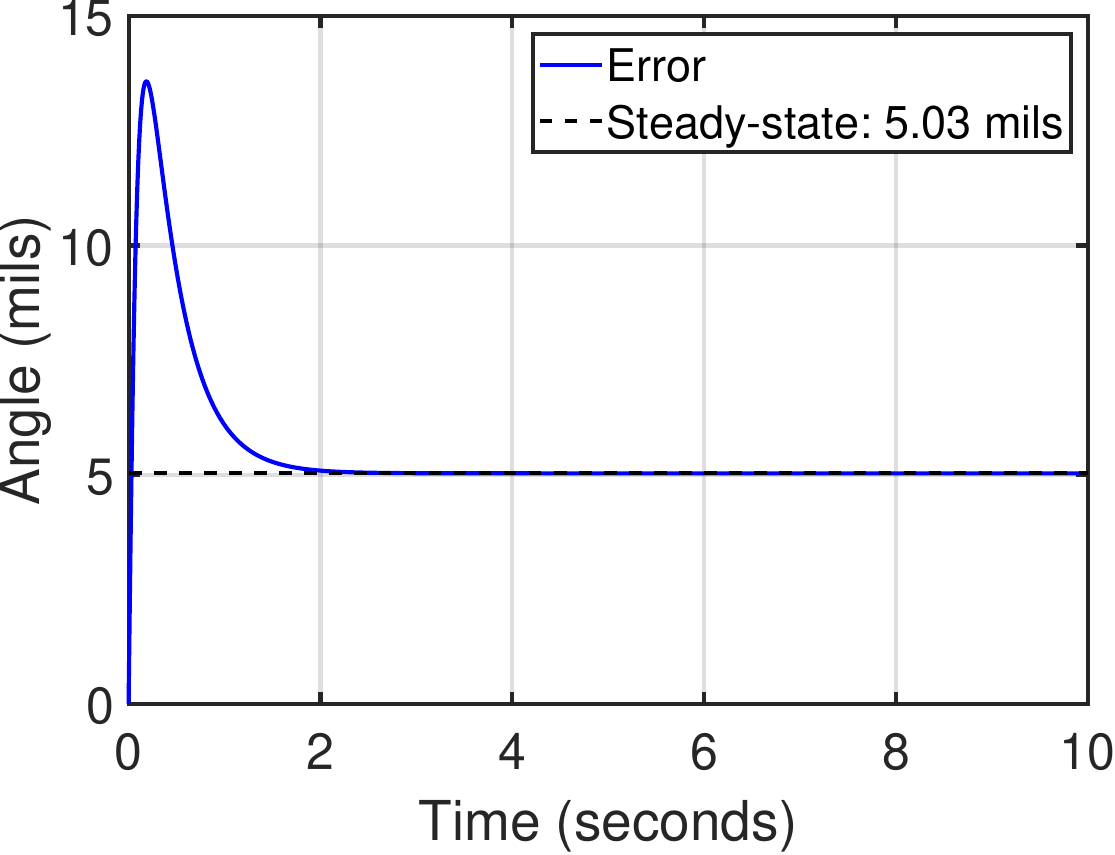}
		\caption{Azimuth}
		\label{pidazimuthleadmterror}
	\end{subfigure}
	\begin{subfigure}{0.32\textwidth}
		\centering
		\includegraphics[width=0.9\linewidth]{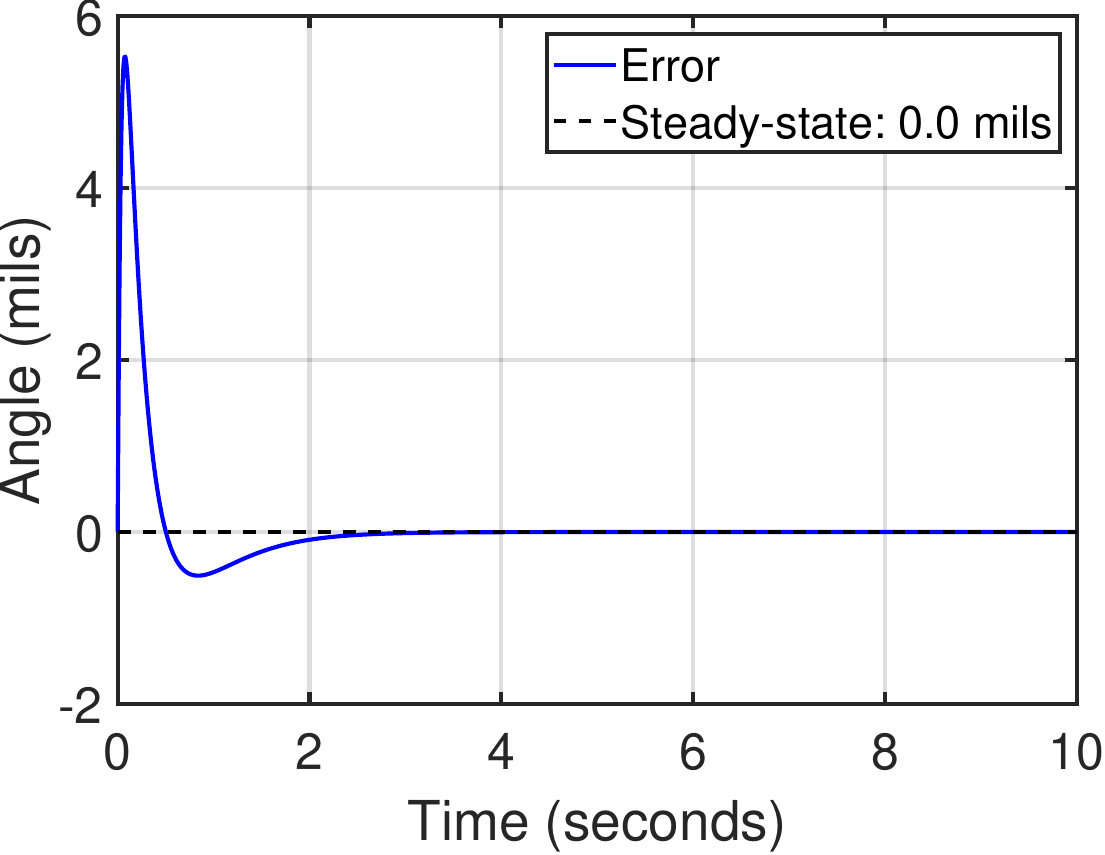}
		\caption{Azimuth}
		\label{pidazimuthpileadmterror}
	\end{subfigure}
	\begin{subfigure}{0.32\textwidth}
		\centering
		\includegraphics[width=0.9\linewidth]{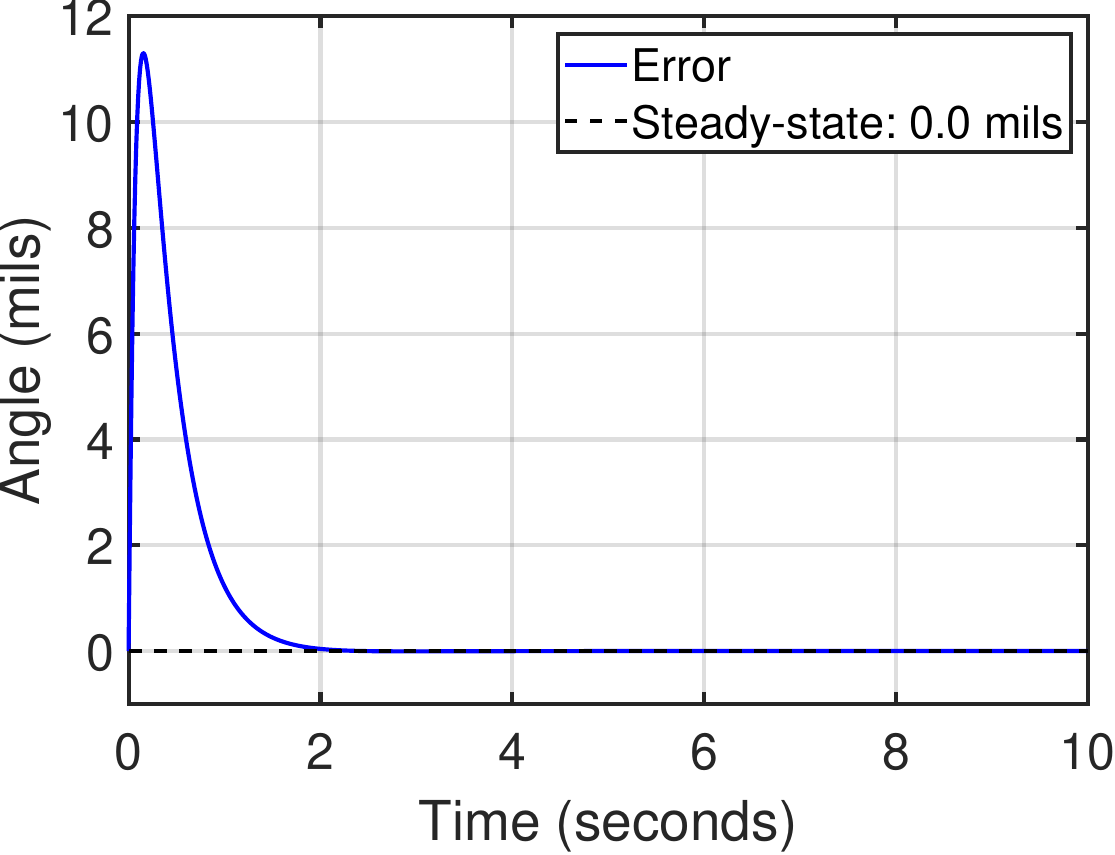}
		\caption{Elevation}
		\label{pidelevationpileadmterror}
	\end{subfigure}
	\caption{Reference tracking output and errors for the azimuth and elevation with different PID controllers. The first column shows the azimuth with lead control, the second column shows the azimuth with PI+lead control, and the third column shows the elevation with PI+lead control.}
\end{figure}

The azimuth output of the gun turret (blue curve) for the lead controller shows a visible offset from the target azimuth position (black-dashed curve) in the inset plot of Figure \ref{pidazimuthleadmtoutput}. Indeed, Figure \ref{pidazimuthleadmterror} shows the tracking error (blue curve) approaching a steady-state value of \SI{5.03}{mils}, indicated by the black-dashed line passing through this point. On the other hand, the azimuth output (blue curve) in Figure \ref{pidazimuthpileadmtoutput} for the PI+Lead controller is shown following the target azimuth position within \num{2} seconds. This is more clearly seen in Figure \ref{pidazimuthpileadmterror} where a plot of the tracking error (blue curve) is approaching a steady-state value of \SI{0}{mils}, indicated by a black-dashed line passing through this point. Likewise, the elevation output (blue curve) of the gun turret with a PI+Lead controller is seen to closely follow the target elevation position (black-dashed curve) in Figure \ref{pidelevationpileadmtoutput} within \num{2} seconds. The tracking error (blue curve) in Figure \ref{pidelevationpileadmterror} is seen approaching a steady-state value of \SI{0}{mil}, indicated by the black-dashed line passing through this point.

The results suggest that integral control is necessary for tracking moving targets due to the positive steady-state error observed in the azimuth with the lead controller in Figure \ref{pidazimuthleadmterror}. The gun turret model with lead control corroborates this outcome as the transfer functions in \eqref{thetatf} and \eqref{alphatf} have one free integrator (one factor of $1/s$). For this type of plant, it can be shown using the Final Value Theorem \citep{Kluever2015Ch8Fvt} that the steady-state errors for the feedback system with a lead controller are non-zero for ramp inputs. Using a PI+lead controller adds an integrator to the closed-loop system, which results in a steady-state error of \SI{0}{mil}.

\section{Conclusions}\label{conclusions}
In this paper, we analyze aiming errors from a  controlled gun turret system given an input target location. A linearized mathematical model of the gun turret is developed and used in controlled turret movement simulations against static and moving targets. We design two different controllers, a PID controller and an MPC controller, to assist in turret movement. The impacts of both errors in estimating the systems' parameters and measurement noise on the aiming accuracy are statistically analyzed. The effects of measurement noise on the aiming errors are modeled and simulation statistics are compared with theoretical results. Preliminary results for tracking moving targets under PID control are presented.

For static targets, the results of our experiments indicate that parameter errors can lead to larger aiming errors on average.
In fact, error in the moment of inertia has a more significant impact on the aiming errors than error in the damping coefficient. These impacts are more evident in the elevation than in the azimuth. However, this could be attributed to sampling the elevation from a smaller range of values. Additionally, the findings support the hypothesis that aiming errors are proportional to target coordinates from stationary position. Accordingly, aiming errors are smaller for gun turret movements over shorter distances than over longer distances. The results also agree with the theory that aiming errors approach zero the longer firing one waits to fire. As a consequence, firing before the settling time of the controller leads to less accuracy, while better accuracy is achieved the longer one waits to fire as expected; in fact, each second one waits to fire results in an increase in accuracy by almost two orders of magnitude. However, this conclusion hinges on the assumption of no time constraints for turret movement, which may not be as useful for modern combat.

In the presence of measurement noise, it is observed that the aiming error distribution mean and standard deviation will match those of the noise the longer one waits to fire. Furthermore, if the measurement noise is normally distributed, the results indicate that aiming errors in steady-state are normally distributed. This is also based on the linear system assumption and agrees with theoretical estimates of the aiming error distribution.

Lastly, we have shown that integral control is needed for tracking moving targets since aiming errors of the gun turret system under lead control are non-zero.

For future work, we will extend this research to include the effects of projectile motion in evaluating the aiming error. This is not considered in the current study. Additionally, we intend to extend the current work to examine the impacts of AI object detection models on the aiming error of the controlled gun turret system. This analysis may include moving target scenarios in addition to static targets.

\section*{Acknowledgment}
The authors would like to gratefully acknowledge Patrick Debroux of the DEVCOM Analysis Center (DAC) for his support and guidance on this work.

\section*{Author Statement}
This research was sponsored by the DEVCOM Analysis Center and was accomplished under Contract Number W911QX--23--D0002. The views and conclusions contained in this document are those of the authors and should not be interpreted as representing the official policies, either expressed or implied, of the Army Research Office or the U.S. Government. The U.S. Government is authorized to reproduce and distribute reprints for Government purposes notwithstanding any copyright notation herein.

\bibliography{./References}

\appendix
\section{Appendix}\label{appendix}

\subsection{Equations of Motion}
\label{eqnsofmotion}

\begin{figure}
    \centering
    \includegraphics[scale=0.5]{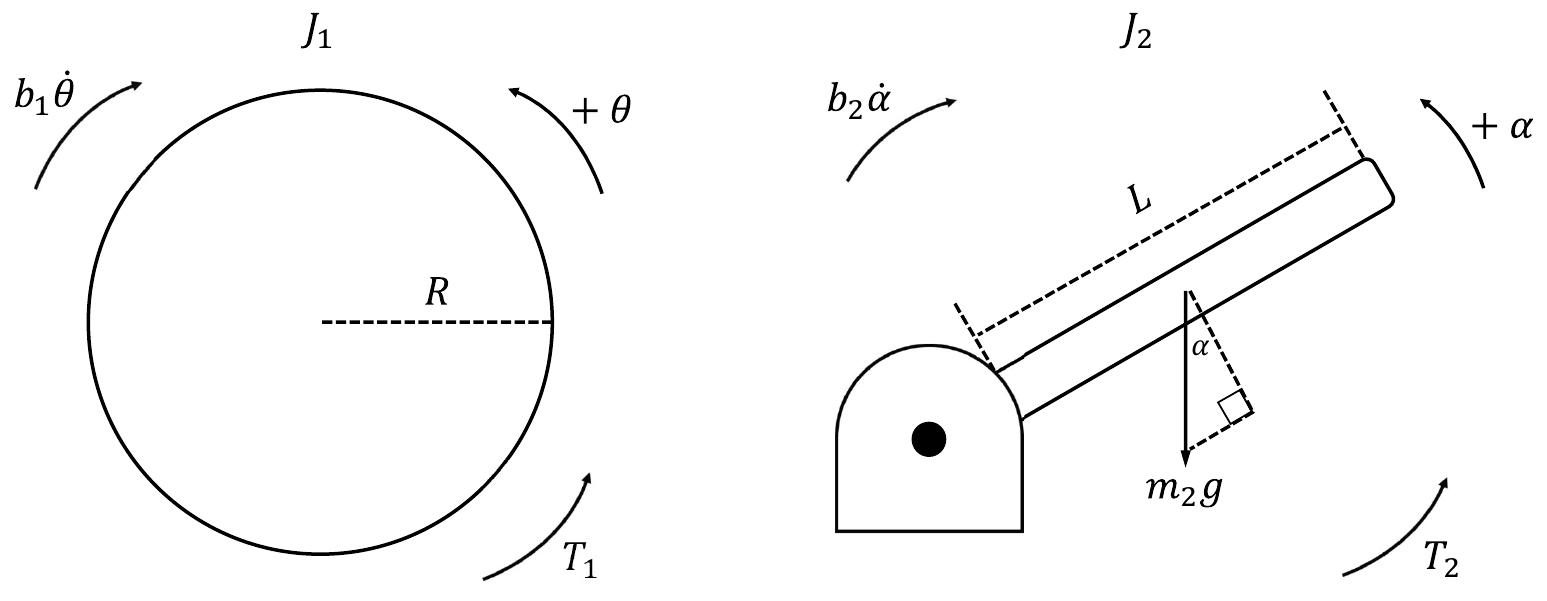}
    \caption{Free body diagram of the platform and gun barrel.}
    \label{free_body_diagram}
\end{figure}

We derive the equations of motion for the gun turret using Newton's second law for angular motion, which is
\begin{equation}
    \sum_i M_{i,O} = J\dot{\omega},
    \label{nslrotation}
\end{equation}
where $M_{i,O}$ is a moment about the point $O$ in a body with moment of inertia $J$ and $\omega$ is the angular velocity. We appeal to the free body diagram in Figure \ref{free_body_diagram} that illustrates the forces and moments acting on the system and the sign convention. Although not shown in the diagram, there is a torque acting on the gun barrel in the clockwise direction due to the weight, which has a moment arm of $L/2$. By summing the moments and applying \eqref{nslrotation}, we obtain the following two equations:

    \begin{align}
        J_1\ddot{\theta}& = -b_1\dot{\theta} + T_1 
        \label{thetaeqnofmotionderiv}\\
        J_2\ddot{\alpha}& = -b_2\dot{\alpha} - \frac{1}{2}m_2gL\cos\alpha + T_2,
        \label{alphaeqnofmotionderiv}
    \end{align}   
Moving all terms with derivatives to the left-hand side of \eqref{thetaeqnofmotionderiv} and \eqref{alphaeqnofmotionderiv} results in equations \eqref{thetaeqnofmotion} and \eqref{alphaeqnofmotion} for the gun turret outputs $\theta$ and $\alpha$. 

\subsection{Linear State-Space Model}
\label{linearssmodelderivation}

We derive the state space model of the gun turret to use in the design of the MPC controller as follows. As a reminder, the states are $x_1 = \theta$, $x_2 = \dot{\theta}$, $x_3 = \alpha$, and $x_4 = \dot{\alpha}$, while the inputs are $u_1 = T_1$ and $u_2 = T_2$, and the outputs are $y_1 = x_1$ and $y_2 = x_2$. Putting the variable definitions into the equations of motion \eqref{thetaeqnofmotionderiv} and \eqref{alphaeqnofmotionderiv} leads to the following system of first-order differential equations:
\begin{align}
    \dot{x}_1 &= x_2 \label{x1doteqn}\\
    \dot{x}_2 &= -\frac{b_1}{J_1}x_2 + \frac{u_1}{J_1} \label{x2doteqn}\\
    \dot{x}_3 &= x_4 \label{x3doteqn}\\
    \dot{x}_4 &= -\frac{b_2}{J_2}x_4
    -\frac{mgL}{2J_2}\cos x_3 + \frac{u_2}{J_2} \label{x4doteqn}\\
    y_1 &= x_1 \label{y1eqn}\\
    y_2 &= x_3.
    \label{y2eqn}
\end{align}

This is a nonlinear system of equations of the form
\begin{align} 
    \dot{\mathbf{x}}(t) &= \mathbf{f}(\mathbf{x}(t),\mathbf{u}(t)) \label{xdotfnc}\\
    \mathbf{y}(t) &= \mathbf{g}(\mathbf{x}(t),\mathbf{u}(t)),
    \label{yfnc}
\end{align}
where $\mathbf{x}(t)$ is a $4x1$ vector of state variables and $\mathbf{u}(t)$ is a $2x1$ vector of inputs. The function $\mathbf{f}(\mathbf{x}(t),\mathbf{u}(t))$ is a $4x1$ vector whose components are the right-hand sides of equations \eqref{x1doteqn}-\eqref{x4doteqn} and the function $\mathbf{g}(\mathbf{x}(t),\mathbf{u}(t))$ is a $2x1$ vector whose components are the right-hand sides of equations \eqref{y1eqn} and \eqref{y2eqn}. 

To obtain a state-space representation, we assume a static equilibrium point exists at the nominal input $\mathbf{u}_0 = [0,mgL/2]^{\mathrm{T}}$. This means that $\dot{\mathbf{x}} = \mathbf{0}$. We then set the left-hand sides of equations \eqref{x1doteqn}-\eqref{x4doteqn} equal to zero and solve the system. This produces the equilibrium point $\mathbf{x}_0 = [0,0,0,0]^{\mathrm{T}}$. 

Next, we calculate the Jacobian matrices
\begin{align} 
\mathbf{A} = \frac{\partial \mathbf{f}}{\partial\mathbf{x}}\Big|_{\mathbf{x}_0}, &\quad \mathbf{B} = \frac{\partial\mathbf{f}}{\partial\mathbf{u}}\Big|_{\mathbf{x}_0} 
\label{statematrices} \\
\mathbf{C} = \frac{\partial \mathbf{g}}{\partial\mathbf{x}}\Big|_{\mathbf{x}_0}, &\quad \mathbf{D} = \frac{\partial \mathbf{g}}{\partial\mathbf{u}}\Big|_{\mathbf{x}_0}. 
\label{outputmatrices}
\end{align}
As an example, we will show the calculation for $\mathbf{A}$, which is a $4\times 4$ matrix with real entries. From equations \eqref{x1doteqn}-\eqref{x4doteqn}, it follows that
\begin{equation}
\frac{\partial f_1}{\partial x_1} = 0, \quad \frac{\partial f_1}{\partial x_2} = 1, \quad
\frac{\partial f_1}{\partial x_3} = 0, \quad
\frac{\partial f_1}{\partial x_4} = 0 \label{df1dx}
\end{equation}
\begin{equation} 
\frac{\partial f_2}{\partial x_1} = 0, \quad  \frac{\partial f_2}{\partial x_2} = -\frac{b_1}{J_1}, \quad \frac{\partial f_2}{\partial x_3} = 0, \quad \frac{\partial f_2}{\partial x_4} = 0 \label{df2dx}
\end{equation}
\begin{equation}
\frac{\partial f_3}{\partial x_1} = 0, \quad \frac{\partial f_3}{\partial x_2} = 0, \quad \frac{\partial f_3}{\partial x_3} = 0, \quad \frac{\partial f_3}{\partial x_4} = 1, \label{df3dx}
\end{equation}
\begin{equation}
\frac{\partial f_4}{\partial x_1} = 0, \quad\frac{\partial f_4}{\partial x_2} = 0, \quad \frac{\partial f_4}{\partial x_3} = \frac{mgL}{2J_2}\sin x_3, \quad\frac{\partial f_4}{\partial x_4} = -\frac{b_2}{J_2} \label{df4dx}.
\end{equation}
The partial derivatives are then evaluated at the equilibrium point to obtain $\mathbf{A}$. This procedure is repeated for $\mathbf{B}$, $\mathbf{C}$, and $\mathbf{D}$. The resulting state space matrices at the end of this procedure are
\begin{equation}
      \mathbf{A} = \begin{bmatrix}
        0 & 1 & 0 & 0 \\
        0 & -\frac{b_1}{J_1} & 0 & 0 \\
        0 & 0 & 0 & 1 \\
        0 & 0 & 0 & -\frac{b_2}{J_2}
    \end{bmatrix}
    , \quad
    \mathbf{B}
    =
    \begin{bmatrix}
     0 & 0 \\
     \frac{1}{J_1} & 0 \\
     0 & 0 \\
     0 & \frac{1}{J_2}
    \end{bmatrix}
   \label{stateInputMatrices}
\end{equation}
\begin{equation}
    \mathbf{C}
    = 
    \begin{bmatrix}
    1 & 0 & 0 & 0 \\
    0 & 0 & 1 & 0
    \end{bmatrix}
    , \quad
    \mathbf{D} =
    \begin{bmatrix}
    0 & 0 \\
    0 & 0
    \end{bmatrix}
    \label{outputMatrices}
\end{equation}

\subsection{Lead Controller Design}
\label{leaddesign}

Given a desired $\omega_{gc}$ and PM:
\begin{enumerate}
    \item Calculate the magnitude and phase of the plant at $\omega_{gc}$: $|G(j\omega_{gc})|$, $\angle G(j\omega_{gc})$.
    \item Calculate the amount of phase to be added:
    \begin{equation*}
        \phi_{\mathrm{add}} = \mathrm{PM} - 180^{\circ} - \angle G(j\omega_{gc}).
    \end{equation*}
    \item Calculate $\gamma$:
    \begin{equation*}
        \gamma = \frac{1-\sin\phi_{\mathrm{add}}}{1+\sin\phi_{\mathrm{add}}}.
        \label{alpha}
    \end{equation*}
    \item The phase is added at $\omega_{gc} = 1/\sqrt{\gamma}T_D$. Calculate $T_D$ to ensure the phase is added at $\omega_{gc}$:
    \begin{equation*}
        T_D = \frac{1}{\sqrt{\gamma}\omega_{gc}}.
    \end{equation*}
    \item Calculate $K_p$ to make $|C(j\omega_{gc})G(j\omega_{gc})|$ unity:
    \begin{equation*}
        K_P = \frac{\sqrt{\gamma}}{|G(j\omega_{gc})|}.
        \label{kpcondition}
    \end{equation*}
    \item Check the response and return to step 1 if further adjustment is necessary.
\end{enumerate}

\subsection{PI+lead Controller Design}
\label{pileaddesign}

Given a desired $\omega_{gc}$ and PM:
\begin{enumerate}
    \item Calculate the magnitude and phase of the plant at $\omega_c$: $|G(j\omega_{gc})|$, $\angle G(j\omega_{gc})$.
    \item Calculate the amount of phase to be added:
    \begin{equation*}
        \phi_{\mathrm{add}} = \mathrm{PM} - 180^{\circ} - \angle G(j\omega_{gc}) + 6^{\circ}.
    \end{equation*}
    The extra $6^{\circ}$ accounts for the reduction in phase from PI control.
    \item Calculate $\gamma$:
    \begin{equation*}
        \gamma = \frac{1-\sin\phi_{\mathrm{add}}}{1+\sin\phi_{\mathrm{add}}}.
    \end{equation*}
    \item The phase is added at $\omega_{gc} = 1/\sqrt{\gamma}T_D$. Calculate $T_D$ to ensure the phase is added at $\omega_{gc}$:
    \begin{equation*}
        T_D = \frac{1}{\sqrt{\gamma}\omega_{gc}}.
    \end{equation*}
    \item Calculate $K_p$ to make $|C(j\omega_{gc})G(j\omega_{gc})|$ unity:
    \begin{equation*}
        K_P = \frac{\sqrt{\gamma}}{|G(j\omega_{gc})|}.
    \end{equation*}
    \item Make the zero $1/T_I$ a decade below $\omega_{gc}$:
    \begin{equation*}
        T_I = \frac{10}{\omega_{gc}}.
    \end{equation*}
    \item Check the response and return to step 1 if further adjustment is necessary.
\end{enumerate}
 
\subsection{Error Response Standard Deviation Due to White Noise}
\label{errorstdevderivation}

In the frequency domain, the variance of a white noise signal $u$ \citep{Bendat2010RandomData} is shown in the following:
\begin{equation*}
    \sigma_u^2 = \frac{1}{2\pi}\int_{-\omega_N}^{\omega_N}S_u(\omega)\,d\omega,
\end{equation*}
where $S_u(\omega)$ is the spectral density of $u$ and $\omega_N$ is the Nyquist frequency, or half the sampling frequency in rad/sec. The sampling frequency is $f_s=1/h$, where $h$ is the sample time in units of seconds. Since $u$ is white noise, by definition $S_u(\omega)$ is constant and, therefore, can be taken out of the integral. Using the identity $\omega_N = \pi f_s$, it then follows that
\begin{align*}
    \sigma_u^2 &= \frac{1}{2\pi}S_u(\omega)\int_{-\omega_N}^{\omega_N}\,d\omega \\
    &= \frac{\omega_N}{\pi}S_u(\omega) \\
    &= f_s S_u(\omega).
\end{align*}
We then obtain the identity
\begin{equation}
    S_u(\omega) = \frac{\sigma_u^2}{f_s}.
    \label{whitenoisesd}
\end{equation}

The variance of the error response $y$ to the white noise signal $u$ in the frequency domain is
\begin{equation}
    \sigma_y^2 = \frac{1}{2\pi}\int_{-\omega_N}^{\omega_N}S_y(\omega)\,d\omega,
\end{equation}
where the spectral density of the error response is $S_y(\omega) = |G(e^{j\omega h})|^2S_u(\omega)$. Here $G(z)$, with $z = e^{sh}$ for a complex number $s$, is the discrete-time transfer function between the input $u$ and output $y$.

Using \eqref{whitenoisesd}, the variance of the error becomes
\begin{align*}
    \sigma_y^2 &= \frac{1}{2\pi}\int_{-\omega_N}^{\omega_N} |G(e^{j\omega h})|^2S_u(\omega)\,d\omega \\
    &= \frac{1}{2\pi}\int_{-\omega_N}^{\omega_N}|G(e^{j\omega h})|^2\cdot\frac{\sigma^2}{f_s}\,d\omega \\
    &= \frac{1}{2\pi}\int_{-\omega_N}^{\omega_N}|G(e^{j\omega h})|^2\,d\omega\cdot\frac{\sigma^2}{f_s}\\
    &= \Vert G\Vert_2^2\frac{\sigma_u^2}{f_s},
\end{align*}
where $\Vert G \Vert_2$ is the 2-norm defined in \eqref{2norm}.

Taking the square-root on both sides gives the standard deviation of the error response 
$$\sigma_y = \Vert G\Vert_2\frac{\sigma_u}{\sqrt{f_s}}.$$

\end{document}